\documentclass[12pt]{article}
\usepackage{latexsym,amssymb,amsmath}
\usepackage{amsthm, amstext}
\usepackage{array, amsfonts, mathrsfs}
\usepackage{mathrsfs}
\usepackage[dvips]{graphicx,color}
\usepackage{epsf}

\newtheorem{Theorem}{Theorem}
\newtheorem{Lemma}{Lemma}

\newtheorem{Remark}{Remark}

\date{}
\begin{document}

\title{Eikonal algebra on a graph of simple structure}
\author{M.I.Belishev\thanks {Saint-Petersburg Department of the Steklov Mathematical Institute, RAS,
                 belishev@pdmi.ras.ru;  Saint-Petersburg State University, 7/9 Universitetskaya nab.,
                 St. Petersburg, 199034, Russia, m.belishev@spbu.ru. Supported by the RFBR grant
                 18-01-00269.}
        A.V.Kaplun\thanks{Saint-Petersburg State University, 7/9 Universitetskaya nab.,
                 St. Petersburg, 199034, Russia, alex.v.kaplun@gmail.com. Supported by the RFBR grant
                 18-01-00269.}}
\maketitle

\begin{abstract}
An eikonal algebra ${\mathfrak E}(\Omega)$ is a C*-algebra related
to a metric graph $\Omega$. It is determined by trajectories and
reachable sets of a dynamical system associated with the graph.
The system describes the waves, which are initiated by boundary
sources (controls) and propagate into the graph with finite
velocity. Motivation and interest to eikonal al\-geb\-ras comes
from the inverse problem of reconstruction of the graph via its
dynamical and/or spectral boundary data. Algebra ${\mathfrak
E}(\Omega)$ is determined by these data. In the mean time, its
structure and algebraic invariants (irreducible representations)
are connected with topology of $\Omega$. We demonstrate such
connections and study ${\mathfrak E}(\Omega)$ by the example of
$\Omega$ of a simple structure. Hopefully, in future, these
connections will provide an approach to reconstruction.
\end{abstract}

\section{Introduction}\label{sec Introduction}

\subsubsection*{About the paper}
Eikonal algebras appear in the framework of algebraic version of
the boundary control method (BC-method), which is an approach to
inverse problems based on their rela\-ti\-ons to control and
system theory \cite{B Obzor IP 07}. These algebras are used for
re\-con\-s\-t\-ruction of Riemannian manifolds via dyna\-mi\-cal
and/or spectral boundary inverse data. Namely, these data
determine the relevant eikonal algebra, which is a {\it
commutative} C*-algebra, whereas its spectrum (a set of
irreducible representations) provides an isometric copy of the
manifold under reconstruction and, thus, solves the problem
\cite{B UCLA 2013, BDemch Geom Phys, BDemchPopov Izvest}.

Applications of the BC-method to inverse problems on graphs are
rep\-re\-se\-n\-ted in \cite{B IP graph 2004, BV JIIPP graph of
strings, BWada 1}. An eikonal algebra associated with a metric
graph is introduced in \cite{BWada 2}. It is a straightforward
analog the algebras, which are successfully used for solving the
above-mentioned reconstruction problems. As was hoped, such an
analog should reconstruct graphs. However, this analog turns out
to be much more difficult for the use and study. The main reason
is that the eikonal algebra on a graph is {\it noncommutative}. In
\cite{BWada 2} some general features of its structure are
revealed: it is represented as a sum of the so-called
`block-algebras'. This structure is connected with geometry of the
graph but the connection is of unclear and implicit character.
Moreover, the block-algebras are also of rather complicated subtle
structure.

However, we hope for availability of the eikonal algebra on graphs
and its future role in reconstruction problem. Moreover, it is of
certain independent interest as a C*-algebra associated with a
concrete important inverse problem of mathematical physics. By
this paper we start its systematic study and begin with a simple
example. Our goal is to analyze this example in detail.

\subsubsection*{Contents}
\noindent$\bullet$\,\,\,In section \ref{sec Waves on graph} a
hyperbolic dynamical system associated with a metric graph is
introduced. The system describes the waves, which are initiated by
the sources (controls) acting from the boundary vertices and
propagate into the graph with the unit velocity. The system is
endowed with the control theory attributes: outer and inner
spaces, and operators. The waves constitute the reachable sets
(subspaces) and determine the corresponding projections on them.
An {\it eikonal} is defined as an operator integral composed of
these projections. It acts in the inner space and is determined by
a single boundary vertex. The eikonals corresponding to a set of
boundary vertices generate an {\it eikonal algebra}  ${\mathfrak
E}^T_\Sigma$. It is an operator algebra, which is a key object of
the paper.
\smallskip

\noindent$\bullet$\,\,\,Section \ref{sec Representation of waves}
provides the instruments for analyzing the structure of
${\mathfrak E}^T_\Sigma$. The main role is played by a
parametrization, which represents the waves as elements of the
spaces $L_2([a,b]; {\mathbb R}^m)$ and eikonals as operators
multiplying elements by the matrix-valued functions of the class
$C([a,b]; {\mathbb M}^m)$.
\smallskip

\noindent$\bullet$\,\,\,Section \ref{sec Agebra Eik} contains some
general facts on C*-algebras and, in particular, the matrix
algebras being in the use. We introduce the so-called {\it
block-algebras}, which play the role of the building blocks
constituting ${\mathfrak E}^T_\Sigma$.
\smallskip

\noindent$\bullet$\,\,\,Section \ref{sec Simple graph} is devoted
to analysis of the eikonal algebra of a simple graph. The graph is
a 3-star: it consists of three edges emanating from a single inner
vertex, contains three boundary vertices and is controlled from
two of them. The edges are of the different lengths. The
corresponding dynamical system, which describes the wave
propagation, is considered at the finite time interval $[0,T]$.
The waves propagate from the boundary vertices with the speed $1$
and gradually fill the graph. Respectively, the algebra
${\mathfrak E}^T_\Sigma$ changes as time $T$ goes on. The
evolution of its structure is of our main interest: we analyze it
in detail.

\subsubsection*{Comments}
$\bullet$\,\,\,The bulk of the subject matter of sections \ref{sec
Waves on graph} and \ref{sec Representation of waves} is the same
as in \cite{BWada 2}. We just repeat the basic notions and facts
from \cite{BWada 2} to make the paper appropriate for independent
reading. A new object appears in subsection `More hydras', where
we introduce the so-called {\it efficient hydra}. The novelty
provides more natural and convenient partition of the graph. As a
result, one gets more transparent description of the structure of
${\mathfrak E}^T_\Sigma$.
\smallskip

\noindent$\bullet$\,\,\,The literature on inverse problems on
graphs is hardly observable. We refer the reader to the papers by
S.Avdonin, P.Kuchment, P.Kurasov, V.Yurko and others. Eikonal
algebras are dedicated to solving inverse problems \cite{B UCLA
2013,BDemch Geom Phys,BDemchPopov Izvest}. However, in the given
paper we do not solve inverse problems but study an algebra
closely related to them.  A prospective goal is to recover a graph
via its boundary inverse data by the use of ${\mathfrak
E}^T_\Sigma$.
\smallskip

\noindent$\bullet$\,\,\,The authors are extremely grateful to
I.V.Kubyshkin for kind assistance in computer graphics.

\section{Waves on graph}\label{sec Waves on graph}
\subsubsection*{A graph}
Basically, our results are meaningful and valid for arbitrary
metric graphs \cite{BWada 1, BWada 2}. However, for the sake of
simplicity, we deal with the following specific case.
\smallskip

A graph $\Omega=E\cup W$ is a connected set in ${\mathbb R}^3$
\footnote{So, we don't need the graph to be planar.}, which
consists of the {\it edges} $E=\{e_i\}_{i=1}^L$ and {\it vertices}
$W=\{w_j\}_{j=1}^M$. Each edge is a finite open interval of the
straight line: $e_i=\{x_i+s\omega_i\,|\,\,a_i<s<b_i\}$, where
$x_i,\omega_i\in {\mathbb R}^3,
\,|\omega_i|=1;\,\,a_i,b_i\in\mathbb R$ provided $e_i\cap
e_{i'}=\emptyset$. The vertices $w_j\in{\mathbb R}^3$ are the
endpoints of the edges. We say $w\in W$ to be incident to $e\in E$
(and vice versa) and write $w\prec e$ if $w\in\overline e$ (the
closure in ${\mathbb R}^3$). The number $\mu(w)\geqslant 1$ of the
edges incident to the given $w\in W$ is called a valency of $w$.

In what follows we deal with the graphs obeying $\mu(w)\not=2$, so
that $W=V\cup\Gamma$, where $\Gamma=\{w\in W\,|\,\,\mu(w)=1\}$ and
$V=\{w\in W\,|\,\,\mu(w)\geqslant 3\}$ are the sets of the {\it
boundary} vertices and {\it inner} vertices respectively. Also, we
assume $\Gamma\not=\emptyset$, rename the boundary vertices by
$\{\gamma_1,\dots,\gamma_N\}$ and say $\Gamma$ to be the boundary
of $\Omega$. The inner points are denoted by
$\{v_1,\dots,v_{M-N}\}=V$.

Graph $\Omega$ is endowed with an intrinsic metric $\tau(x,x')$
induced by the ${\mathbb R}^3$-metric and defined as the length of
the shortest path in $\Omega$, which connects $x$ and $y$. In
particular, for $x,x'\in\overline e$ one has $\tau(x,x')=|x-x'|$.
For a subset $A\subset\Omega$, the set
 \begin{equation*}
\Omega^r[A]:=\{x\in\Omega\,|\,\,\tau(x,A)<r\}\,\,\,(r>0)
 \end{equation*}
is its metric neighborhood of radius $r$. Thus, $\Omega$ is a
compact connected metric space.

\subsubsection*{Derivatives, spaces, operators}
In the paper all the functions, function classes and spaces are
real.
\smallskip

\noindent$\bullet$\,\,\,Let $e=\{x(s)=\tilde
x+s\omega\,|\,\,a<s<b\}$ be a (parametrized) edge, $y$ a function
on $\Omega$. For a point $x=x(s)\in e$ one defines a derivative
along the edge by
 \begin{equation*}
\partial_e y(x)\,:=\,\lim \limits_
       {\delta \to 0} \frac{y(x(s+\delta))-y(x(s))}{\delta}\,.
 \end{equation*}
Such a derivative depends on the parametrization up to the sign.
In the mean time, the second derivative $\partial^2_e y(x)$ is
invariant.

For a vertex $w$ and edge $e$ provided $w\prec e$, one defines an
{\it outward} derivative
 \begin{equation*}
\partial^+_e y(w)\,:=\,\lim \limits_
       {e \ni \,x \to \,w} \frac{y(x)-y(w)}{\tau(x,w)}\,.
 \end{equation*}
For an interior vertex $v \in V$,  an {\it outward flow}
 \begin{equation*}
\Pi_v[y]\,:=\, \sum \limits_{e\succ v}
\partial^+_e y(v)
 \end{equation*}
is introduced.
\smallskip

\noindent$\bullet$\,\,\,By $C(\Omega)$ we denote the Banach space
of continuous functions with the norm $\|y\|=\underset{\Omega}{\rm
sup\,}|y(\cdot)|$.

Introduce a Hilbert space ${\mathscr H}=L_2(\Omega)$ of functions
with the inner product
 $$
(y,u)_{\mathscr H}:=\int _\Omega y u\,d\tau = \sum_{e \in
E}\,\,\int _e y u\,d\tau\,,
 $$
where $d\tau$ is the length element on $\Omega$.

A function $y\in C(\Omega)$ is assigned to a class $H^2(\Omega)$
if $y(x(\cdot))$ belongs to the Sobolev class $H^2(a,b)$ for each
edge $e=\{x(s)=\tilde x+s\omega\,|\,\,a<s<b\}$. Also, we define
the {\it Kirchhoff class}
 \begin{align}\label{Eq Kirchhoff class}
{\mathscr K} \,:= \,\{y \in
{H}^2(\Omega)\,|\,\,\,\Pi_v[y]:=0,\,\,\,v \in V \}\,.
 \end{align}

\noindent$\bullet$\,\,\,The {\it Laplace operator}
 $\Delta: {\mathscr  H}\to{\mathscr H},
\,\, {\rm Dom}\, \Delta={\mathscr K}$,
 \begin{equation}\label{Eq Laplace operator}
\left (\Delta y\right )\big|_e\,:=\,\partial^2_e y\,, \qquad e \in
E\,
 \end{equation}
is densely defined and closed.

\subsubsection*{Dynamical system}
\noindent$\bullet$\,\,\,An initial boundary value problem
\begin{align}\label{Eq DS 1}
& u_{tt}-\Delta u=0 && {\rm in}\,\,{\mathscr H},
\,\,\,0<t<T\\
\label{Eq DS 2} &u \in {\mathscr K}  && {\rm
for\,\,all\,}\,t \in [0,T]\\
\label{Eq DS 3} & u|_{t=0}=u_t|_{t=0}=0 && {\rm in}\,\, \Omega\\
\label{Eq DS 4} & u=f && {\rm on \,}\, \Gamma \times [0,T]
\end{align}
is referred to as a {\it dynamical system} associated with the
graph $\Omega$. Here $T >0$ is a final moment;\, $f=f(\gamma ,t)$
is a {\it boundary control}; a solution $u=u^f(x,t)$ describes a
{\it wave} initiated at $\Gamma$ and propagating into $\Omega$. As
is well known, for a $C^2$-smooth (with respect to $t$) control
$f$ vanishing near $t=0$ the problem has a unique classical
solution $u^f$. Later on, the (generalized) solutions for $f\in
L_2(\Gamma\times[0,T])$ will be defined.

By definition (\ref{Eq Kirchhoff class}), the condition (\ref{Eq
DS 2}) yields the {\it Kirchhoff laws}:
 \begin{equation*}\label{Eq Kirchhoff laws}
u(\,\cdot\,,t) \in C(\Omega), \quad \Pi_v[u(\,\cdot\,,t)]=0 \qquad
\text{for all}\,\,\,t \geqslant 0\,\,\, \text{and}\,\, v \in V\,.
 \end{equation*}
By (\ref{Eq Laplace operator}), on each (parametrized) edge $e \in
E$ the function $\tilde u=u^f(x(s),t)$ satisfies the homogeneous
string equation
 \begin{equation}\label{Eq string eqn}
\tilde u_{tt}-\tilde u_{ss}=0 \qquad {\rm in\,}\,\,(a,b)\times
(0,T).
 \end{equation}
Hence, the waves propagate in $\Omega$ with the unit speed.
\smallskip

\noindent$\bullet$\,\,\,A space of controls ${\mathscr F}^T
:=L_2\, (\Gamma \times [0,T])$ with the inner product
 $$
(f,g)_{{\mathscr F}^T} :=\sum\limits_{\gamma \in \Gamma} \,
\int_{0}^T f(\gamma ,t)\,g(\gamma ,t)\,dt
 $$
is called an {\it outer space} of system (\ref{Eq DS 1})--(\ref{Eq
DS 4}). It contains the subspaces
 $$
{\mathscr F}^T_\gamma:=\left\{f \in {\mathscr F}^T\,|\,\, {\rm
supp\,}f \subset \{\gamma\} \times [0,T]\right\}
 $$
of controls, which act from single boundary vertices $\gamma \in
\Gamma$. Each $f \in {\mathscr F}^T_\gamma$ is of the form
$f(\gamma',t)=\delta_\gamma(\gamma')\varphi(t)$, where For
 \begin{equation*}
\delta_\gamma (\gamma'):= \begin{cases} 0, & \gamma'\not= \gamma
\\1, & \gamma'=\gamma\end{cases}
 \end{equation*}
and $\varphi \in L_2[0,T]$.

For a subset $\Sigma\subseteq\Gamma$ we put
\begin{equation}\label{Eq F=sum F gamma}
{\mathscr F}^T_\Sigma\,:=\,\oplus \sum_{\gamma \in
\Sigma}{\mathscr F}^T_\gamma\,
\end{equation}
and have ${\mathscr F}^T=\oplus \sum_{\gamma \in \Gamma}{\mathscr
F}^T_\gamma$.
\smallskip

\noindent$\bullet$\,\,\,The space ${\mathscr H}$ is an {\it inner
space}; the waves $u^f(\,\cdot\, ,t)$ are time--dependent elements
of ${\mathscr H}$.
The linear set of waves
 \begin{equation*}
{\mathscr U}^s_\gamma:=\left\{u^f(\cdot,s)\,|\,\,f \in {\mathscr
F}^T_\gamma\right\}\,\subset\,{\mathscr H}, \qquad 0\leqslant
s\leqslant T
 \end{equation*}
is called {\it reachable} (from the boundary vertex $\gamma$, at
the moment $t=s$). We say the set
 \begin{equation*}
{\mathscr U}^s_\Sigma:=\left\{u^f(\cdot,s)\,|\,\,f \in {\mathscr
F}^T_\Sigma\right\}={\rm span}\{{\mathscr
U}^s_\gamma\,|\,\,\gamma\in\Sigma\}
 \end{equation*}
(algebraic sum of ${\mathscr U}^s_\gamma$) to be reachable from
$\Sigma$. As will be noticed later in Remark \ref{Rem 1},
${\mathscr U}^s_\gamma$ and ${\mathscr U}^s_\Sigma$ are the
(closed) subspaces in $\mathscr H$. They are increasing as $s$
grows: ${\mathscr U}^s_\Sigma\subset{\mathscr U}^{s'}_\Sigma$ for
$s<s'$.

\subsubsection*{Eikonals}
Here we introduce the algebra which is the main subject of the
paper. By ${\mathfrak B}({\mathscr H})$ we denote the algebra of
bounded operators acting in the inner space.

Let $P^s_\gamma$ be the (orthogonal) projection in ${\mathscr H}$
onto ${\mathscr U}^s_\gamma$. The operator
$E^T_\gamma\in{\mathfrak B}({\mathscr H})$ of the form
 \begin{equation*}
E^T_\gamma\,:=\,\int^T_0 s\,dP^s_\gamma
 \end{equation*}
is called an {\it eikonal} corresponding to the vertex $\gamma$.

For a Banach algebra $\mathfrak B$ and a subset $A\subset\mathfrak
B$ by $\vee A$ we denote the minimal closed (sub)algebra in
$\mathfrak B$ which contains $A$. The algebra
 \begin{equation}\label{Eq def alg eik}
{\mathfrak
E}^T_\Sigma\,:=\,\vee\{E^T_\gamma\,|\,\,\gamma\in\Sigma\}
 \end{equation}
generated by eikonals is called the {\it eikonal algebra}
\cite{BWada 2}. Our general goal is to study its structure.

\section{Representation of waves and eikonals}\label{sec Representation of waves}
Here we derive a relevant representation for elements of
${\mathfrak E}^T_\Sigma$.

\subsubsection*{Generalized solutions}
\noindent$\bullet$\,\,\,Consider the system (\ref{Eq DS
1})--(\ref{Eq DS 4}) with $T=\infty$. Let $\delta(t)$ be the Dirac
delta-function.

Fix a boundary vertex $\gamma$. Taking the control
$f(\gamma',t)=\delta_\gamma (\gamma')\delta(t)$, one can define
the (generalized) solution $u^{\delta_\gamma \delta}$ to (\ref{Eq
DS 1})--(\ref{Eq DS 4}). A possible way is to use a smooth
regularizations $\delta^\varepsilon(t) \underset{\varepsilon \to
0}\to \delta(t)$ and then understand $u^{\delta_\gamma \delta}$ as
a relevant limit of the classical solutions $u^{\delta_\gamma
\delta^\varepsilon}$ as $\varepsilon \to 0$. Such a limit turns
out to be a space-time distribution on $\Omega \times [0,T]$ of
the class $C\left((0,T); H^{-1}(\Omega)\right)$: see, e.g.,
\cite{ContrGraph}. The distribution $u^{\delta_\gamma \delta}$ is
called a {\it fundamental} solution to (\ref{Eq DS 1})--(\ref{Eq
DS 4}) corresponding to the given $\gamma$. It describes the wave
initiated by in\-stan\-taneous source supported at $\gamma$. Let
us consider its properties in more detail. It is convenient to use
the formal rule, which may be specified as `dynamics of particles'
\cite{BWada 2}. By a {\it measure} is meant a linear continuous
functional on the space $C(\Omega)$. A Dirac measure $\delta_x$
acts by $\langle\delta_{x_0},y\rangle=y(x_0)$. The measures
$a\delta_{x(t)}$ are said to be the {\it particles}, the factors
$a\in\mathbb R$ being called {\it amplitudes}.

The rule is the following.
\smallskip

\noindent$1.$\,\,\, Each particle $a\delta_{x(t)}$ moves along an
edge with velocity $1$ in one of two possible directions, so that
$|\dot x(t)|=1$ holds as $x(t)\in e$.
\smallskip

\noindent$2.$\,\,\,Particles move independently, they do not
interact. If by the moment $t$ there are a few particles
$a_1\delta_{x(t)},\dots,a_p\delta_{x(t)}$ supported at the point
$x(t)\in\Omega\backslash\Gamma$, they are identified with the
single particle $[a_1+\dots + a_p]\delta_{x(t)}$.
\smallskip

\noindent$3.$\,\,\,The boundary of the graph reflects particles.
As soon as a particle $a\delta_{x(t)}$ reaches a
$\gamma\in\Gamma$, it instantly reverses its direction and changes
the amplitude from $a$ for $-a$.
\smallskip

\noindent$4.$\,\,\,Moving along the edge $e$ and passing through
an inner vertex $v\prec e$, the particle $a\delta_{x(t)}$ splits
into $\mu(v)$ particles: one reflected and $\mu(v)-1$
trans\-mit\-ted. The reflected particle moves along $e$ in the
opposite direction and is of the amplitude
$\frac{2-\mu(v)}{\mu(v)}a$. Each of the transmitted particles
moves along the single (incident to $v$) edge away from $v$ and
has the amplitude $\frac{2}{\mu(v)}a$ \footnote{Thus, the total
amplitude is
$\frac{2-\mu(v)}{\mu(v)}\,a+[\mu(v)-1]\frac{2}{\mu(v)}\,a=a$ that
corresponds to the Kirchhoff conservation lows.}.
\smallskip

Accepting such a convention, we can describe the solution
$u^{\delta_\gamma\delta}$ as follows. Recall that $\tau$ is the
distance in $\Omega$.
\smallskip

\noindent$\bf A.$\,\,\,For $0\leqslant t\leqslant\tau(\gamma,V)$,
one has $u^{\delta_\gamma\delta}=\delta_{x(t)}$, where $x(t)$ is
the point of the edge $e\succ\gamma$ provided
$\tau(x(t),\gamma)=t$. Thus, for the `small' times,
$u^{\delta_\gamma\delta}$ is a single particle injected from
$\gamma$ into the graph and moving along $e$ with the unit
velocity.
\smallskip

\noindent$\bf B.$\,\,\,Further evolution for the times
$t>\tau(\gamma,V)$ is governed by the rule $1.- 4.$
\smallskip

As is easy to recognize, such a description is quite
deterministic. So, at any moment $t\geqslant 0$ the solution
$u^{\delta_\gamma\delta}$ is a collection of finite number of
particles moving into $\Omega$.
\smallskip

\noindent$\bullet$\,\,\,The fundamental solution is a space-time
distribution on $\Omega\times\{t\geqslant 0\}$. Owing to its above
described specific structure, the function
 \begin{equation}\label{Eq generalized u^f}
 u^f(x,t)\,:=\,\left[u^{\delta_\gamma\delta}(x,\cdot)\ast
 f\right](t), \qquad x\in\Omega,\,\,\,0\leqslant t\leqslant T
 \end{equation}
(the convolution w.r.t. $t$) is well defined for any boundary
control $f\in{\mathscr F}^T_\gamma$ of the form
$f(\gamma',t)=\delta_\gamma(\gamma')\varphi(t)$ with $\varphi\in
L_2[0,T]$. Moreover, one can show that $u^f\in C([0,T];{\mathscr
H})$ and, if $\varphi$ is $C^2$-smooth and vanishes near $t=0$
then $u^f$ provides the classical solution to (\ref{Eq DS
1})--(\ref{Eq DS 4}).

By the aforesaid, we regard $u^f$ defined by (\ref{Eq generalized
u^f}) as a {\it generalized} solution to (\ref{Eq DS 1})--(\ref{Eq
DS 4}) for $f\in{\mathscr F}^T_\gamma$. Also, in accordance with
(\ref{Eq F=sum F gamma}), for $f \in{\mathscr
F}^T:\,f=\sum_{\gamma\in\Gamma}f_\gamma$ with
$f_\gamma\in{\mathscr F}^T_\gamma$, we put
 \begin{equation}\label{Eq generalized u^f total}
 u^f(x,t)\,=\,\sum_{\gamma\in\Gamma}u^{f_\gamma}(x,t), \qquad x\in\Omega,\,\,\,0\leqslant t\leqslant
 T\,.
 \end{equation}
Later on an efficient representation of the waves $u^f$ will be
provided.
\smallskip

\noindent$\bullet$\,\,\,As was mentioned above (see (\ref{Eq
string eqn})), the waves propagate with the unit speed. As a
consequence, for $f\in {\mathscr F}^T_\gamma$ one has
 \begin{equation}\label{Eq supp u^f f in F gamma}
{\rm
supp\,}u^{f}(\cdot,t)\,\subset\,\overline{\Omega^t[\gamma]},\qquad
t>0\,.
 \end{equation}
By the latter and (\ref{Eq generalized u^f total}), for a subset
$\Sigma\subseteq\Gamma$ we have
 \begin{equation*}
{\rm
supp\,}u^{f}(\cdot,t)\,\subset\,\overline{\Omega^t[\Sigma]}\qquad
{\text{for}}\quad f\in\oplus\sum\limits_{\gamma\in\Sigma}{\mathscr
F}^T_\gamma,\,\,\,t>0\,.
 \end{equation*}
Thus, $\Omega^t[\Sigma]$ is the part of the graph filled by waves,
which move from $\Sigma$, at the moment $t$.
\smallskip

\subsubsection*{Hydra}
\begin{figure}[h!]
\centering \epsfysize=7cm \epsfbox{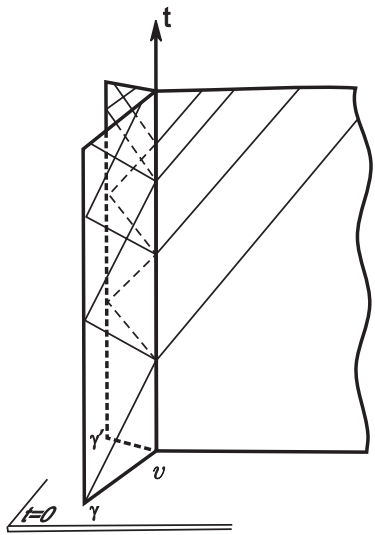} \caption[Fig.1]{The
hydra}\label{The hydra}
 \end{figure}

\noindent$\bullet$\,\,\,Fix a boundary vertex $\gamma$.
Considering the fundamental solution as a space-time distribution,
we int\-ro\-duce the set
 $$
H_\gamma\,:=\,{\rm supp\,u^{\delta_\gamma\delta}}\subset
\Omega\times{\overline{\mathbb R}}_+
 $$
and call it a {\it hydra}\, \cite{BWada 2}.  Thus, the hydra is a
space-time graph formed by trajectories of particles: see Fig.
\ref{The hydra}.
\smallskip

Let
 \begin{align}\label{Eq def pi rho on H gamma}
\notag & \pi: H_\gamma \ni h=(x,t)\mapsto x\in \Omega, \quad
\pi^{-1}(x):=\{h\in H_\gamma\,|\,\,\pi(h)=x\};\\
& \rho: H_\gamma \ni h=(x,t)\mapsto t\in {\overline{\mathbb
R}}_+,\quad \rho^{-1}(t):=\{h\in H_\gamma\,|\,\,\rho(h)=t\}
 \end{align}
be the space-\, and time-projections. On the hydra one defines a
function ({\it amplitude}) $a(\cdot)$ as follows:
\smallskip

\noindent(*)\,\,\,for $h\in H_\gamma$ provided
$\pi(h)=x\in\Omega\backslash\Gamma$ and $\rho(h)=t>0$ we have
$u^{\delta_\gamma \delta}(\cdot,t)=a\delta_x(\cdot)$ and define
$a(h)=a$;
\smallskip

\noindent(**)\,\,\,for $h\in H_\gamma$ provided $\pi(h)\in \Gamma$
and $\rho(h)>0$, we put $a(h)=0$;
\smallskip

\noindent(***)\,\,\,for $h\in H_\gamma$ provided $\pi(h)=\gamma$
and $\rho(h)=0$, we put $a(h)=1$.
\smallskip

So, the amplitude is a piece-wise constant function defined on the
whole $H_\gamma$ and determined by amplitudes of particles: see
Fig. \ref{The amplitude on the hydra}.
 \begin{figure}[h!]
\centering \epsfysize=6cm \epsfbox{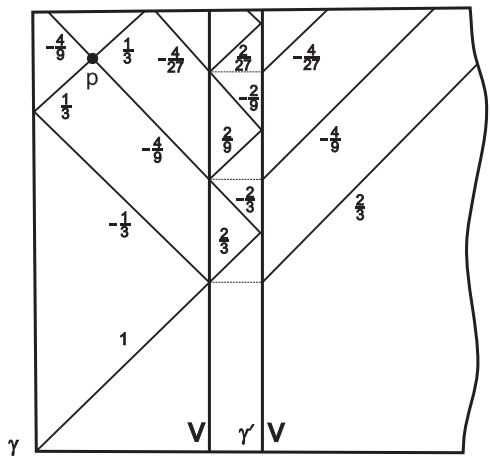}
\caption[Fig.2]{The amplitude on the hydra}\label{The amplitude on
the hydra}
 \end{figure}
As is easy to recognize, this definition is consistent with the
`dynamics of particles' $1.-4.$ In the crossing points $p$, by the
rule $2.$ one has $a(p)=-\frac{4}{9}+\frac{1}{3}=-\frac{1}{9}$.
Also, for $h=(x,t)\in H_\gamma$ we write the amplitude as
$a(x,t)$.
\smallskip

In what follows the basic object is a {\it truncated hydra}
 $$
H^T_\gamma\,:=\,H_\gamma \cap \{\Omega\times[0,T]\}\,.
 $$

\noindent$\bullet$\,\,\,Fix $\gamma\in\Gamma$ and take a control
$f\in {\mathscr
F}^T_\gamma:\,\,\,f(\gamma',t)=\delta_\gamma(\gamma')\varphi(t)$
with $\varphi\in L_2[0,T]$. Let $u^f$ be the wave, i.e., the
(generalized) solution to (\ref{Eq DS 1})--(\ref{Eq DS 4}). As is
shown in \cite{BWada 2}, the representation
 \begin{equation}\label{Eq repres u single gamma}
u^f(x,T)\,=\,\sum\limits_{t\in
\rho(\pi^{-1}(x))}a(x,t)\,\varphi(T-t)\,,\qquad x\in\Omega
 \end{equation}
is valid. In the general case, for $f\in {\mathscr F}^T_\Sigma$:
$\,\,f=\sum_{\gamma\in\Sigma}\delta_\gamma\varphi_\gamma$ with
$\varphi_\gamma\in L_2[0,T]$ one has
 \begin{equation}\label{Eq repres u general}
u^f(x,T)\,=\sum\limits_{\gamma\in\Sigma}\,\sum\limits_{t\in
\rho(\pi^{-1}(x))}a_\gamma(x,t)\,\varphi_\gamma(T-t)\,,\qquad
x\in\Omega\,,
 \end{equation}
where $a_\gamma$ are the amplitudes of the hydras $H^T_\gamma$.
 \begin{Remark}\label{Rem 1}
Representations (\ref{Eq repres u single gamma}) and (\ref{Eq
repres u general}) easily imply that the reachable sets ${\mathscr
U}^s_\gamma$ and ${\mathscr U}^s_\Sigma$ are the (closed)
subspaces in $\mathscr H$.
 \end{Remark}

\subsubsection*{Partition $\Pi$}
Before reading this section, we'd recommend the reader to look
through the paper  \cite{BWada 2}, where the objects introduced
here are described in more detail and provided with pictures.

Recall that $\Omega^T[\gamma]$ is the part of the graph filled at
the final moment $t=T$ by waves moving from $\gamma$\,\,(see
(\ref{Eq supp u^f f in F gamma})). Begin with its special
partition imposed by the structure of hydra $H^T_\gamma$.
\smallskip

\noindent$\bullet$\,\,\,We say the points $h,h'\in H^T_\gamma$ to
be the {\it neighbors} and write $h\sim h'$ if either or both
$\pi(h)=\pi(h')$ and $\rho(h)=\rho(h')$ holds.

We put $h\overset{\gamma}{\sim} h'$ if there are $h'_1,\dots
h'_k\,\in H^T_\gamma$ such that $h\sim h'_1\sim\dots \sim h'_k
\sim h'$, i.e. $h$ and $h'$ are connected via a set of neighbors.
As is clear, $\overset{\gamma}{\sim}$ is an equivalence. The
equivalence class
 $$
{\mathscr L}[h]:=\{h'\in H^T_\gamma\,|\,\,h'\overset{\gamma}{\sim}
h\}
 $$
is called a {\it lattice}. For a subset $B\subset H^T_\gamma$ one
defines the lattice
 $$
{\mathscr L}[B]:=\bigcup\limits_{h\in B}{\mathscr L}[h]\,.
 $$
It is easy to verify that the operation $B\mapsto {\mathscr L}[B]$
possesses the following properties:
 \begin{align*}
& B\subset {\mathscr L}[B];\qquad {\mathscr L}[{\mathscr
L}[B]]={\mathscr L}[B];\qquad {\mathscr L}[B_1\cup B_2]={\mathscr
L}[B_1]\cup{\mathscr L}[B_2];\\
& \pi^{-1}(\pi({\mathscr L}[B]))=\rho^{-1}(\rho({\mathscr
L}[B]))={\mathscr L}[B]\,.
 \end{align*}
By the way, the first three properties show that this operation is
a topological closure.
\smallskip

\noindent$\bullet$\,\,\,With every
$x\in\overline{\Omega^T[\gamma]}\setminus\Gamma$ we associate a
set
 \begin{equation}\label{Eq def Lambda}
\Lambda[x]\,:=\,\pi({\mathscr L[\pi^{-1}(x)]})\subset
\overline{\Omega^T[\gamma]}
 \end{equation}
and name it by a {\it determination set} of the point $x$. Since
$\overset{\gamma}{\sim}$ is an equivalence, the following
alternative holds:
 \begin{equation}\label{Eq Lambda alternative}
\text{for}\,\,\, x\not=x'\,\,\,\text{one\,\,has\, either}\,\,\,
\Lambda[x]=\Lambda[x']\,\,\,\text{or}\,\,\,
\Lambda[x]\cap\Lambda[x']=\emptyset\,.
 \end{equation}

\noindent$\bullet$\,\,\,We say $h\in H_\gamma$ to be a {\it corner
point} if either $\pi(h)\in V\cup\Gamma$ or $h$ is a crossing
point (like $p$ on Fig.2). On the truncated hydra $H^T_\gamma$,
the points of the set $\rho^{-1}(T)$ are also assigned to be
corner points. By ${\rm Corn\,}H^T_\gamma$ we denote the set of
all corner points of the truncated hydra.

The lattice ${\mathscr L}[{\rm Corn\,}H^T_\gamma]$ divides the
hydra into a finite number of the open space-time intervals, the
amplitude $a$ taking a constant value on each interval.

The points, which constitute the set
 \begin{equation}\label{Eq def Theta}
\Theta\,:=\,\pi\left({\mathscr L}[{\rm
Corn\,}H^T_\gamma]\right)\subset \overline{\Omega^T[\gamma]}\,,
 \end{equation}
are called {\it critical}. The critical points divide
$\Omega^T[\gamma]$ into parts. The set
 \begin{equation}\label{Eq def Pi}
\Pi\,:=\,\overline{\Omega^T[\gamma]}\setminus \Theta
 \end{equation}
is a sum of the finite number of open intervals, each interval
belonging to the certain edge $e$. It provides the partition of
$\Omega^T[\gamma]$ consistent with the structure of the hydra
$H^T_\gamma$.
\smallskip

\noindent$\bullet$\,\,\,Let $\omega\subset \Pi$ be a maximal
interval, which does not contain critical points
\footnote{`maximal' means that any bigger interval $\omega'\subset
\Omega^T[\gamma]:\,\omega'\supset\omega$ does contain critical
points}. As is easy to see, the set
 \begin{equation}\label{Eq def Phi}
\Phi\,:=\,\pi\left({\mathscr L}[\pi^{-1}(\omega)]\right)
 \end{equation}
consists of the maximal intervals $\omega_1,\dots,\omega_m$ of the
same length:
 \begin{equation*}
\Phi\,=\,\bigcup\limits_{k=1}^m \omega_k\,, \qquad {\rm
diam\,}\omega_1=\dots={\rm diam\,}\omega_m=:\epsilon_\Phi\,.
 \end{equation*}
We say the intervals $\omega_k$ to be the {\it cells} of the {\it
family} $\Phi$. Comparing the definitions (\ref{Eq def Lambda})
and (\ref{Eq def Phi}), one can easily get the representation
 \begin{equation}\label{Eq Phi= sum Lambda}
\Phi\,=\,\bigcup\limits_{x\in\omega}\Lambda[x]\,,
 \end{equation}
where $\omega$ is any of the cells of $\Phi$.
\smallskip

\noindent$\bullet$\,\,\,Taking another maximal $\omega\subset\Pi$,
which does not belong to the family $\Phi$, one determines another
family consisting of cells, and so on.

As a result, the set $\Pi$ is a collection of disjoint families
$\Phi^1,\dots ,\Phi^J$, each family consisting of disjoint cells:
 \begin{equation}\label{Eq Pi= sum Phi=sum Phi sum omega}
\Pi\,=\,\bigcup\limits_{j=1}^J\Phi^j\,=\,\bigcup\limits_{j=1}^J\bigcup\limits_{k=1}^{m_j}\omega^j_k\,,
 \end{equation}
where $m_j$ is the number of cells in $\Phi^j$. Of course, the
structure (\ref{Eq Pi= sum Phi=sum Phi sum omega}) changes as
$T>0$ varies.
\smallskip

\noindent$\bullet$\,\,\,In parallel to the definition (\ref{Eq def
Lambda}), with every
$x\in\overline{\Omega^T[\gamma]}\setminus\Gamma$ one associates a
set
 \begin{equation}\label{Eq def Ksi}
\Xi[x]\,:=\,\rho({\mathscr L[\pi^{-1}(x)]})\subset [0,T]
 \end{equation}
and verifies that for $x\not=x'$ one has either $\Xi[x]=\Xi[x']$
or $\Xi[x]\cap\Xi[x']=\emptyset$. We put $\Xi[B]:=\cup_{x\in
B}\Xi[x]$.

Let $\Phi=\bigcup_{k=1}^{m_\Phi}\omega_k\subset\Pi$ be a family.
As is easy to see, the set
 \begin{equation*}
\Psi\,:=\,\Xi[\Phi]\,=\,\bigcup_{i=1}^{n_\Phi}\psi_i\subset [0,T]
 \end{equation*}
consists of the time intervals $\psi_i=(t_{i-1},t_i)$ such that
$0\leqslant t_1<t_2\leqslant t_3<t_4\leqslant \dots t_{n_\Phi
-1}<t_{n_\Phi}\leqslant T$, the intervals being of the same length
$t_i-t_{i-1}=\epsilon_\Phi$. We say $\Psi$ also to be a family
consisting of the {\it time cells} $\psi_i$.
\smallskip

In the sequel one makes use of the functions $\tau^i_\Phi: \Phi
\to [0,T]$ defined as follows. For $x\in\Phi$ we put
 \begin{equation}\label{Eq tau i}
\tau^i(x)\,:=\,\psi_i\cap \rho({\mathscr L}[\pi^{-1}(x)])\,,\qquad
i=1,\dots, n_\Phi\,.
 \end{equation}
Since ${\mathscr L}[\pi^{-1}(x)]={\mathscr L}[\pi^{-1}(x_k)]$ for
any $x_k\in\Lambda[x]$, we have $\tau^i(x)=\tau^i(x_k)$. As $x$
varies over the cell $\omega\subset\Phi$, the value $\tau^i(x)$
sweeps the cell $\psi^i\subset\Psi$. Loosely speaking, $\tau^i$ is
a piece-wise linear function on $\Phi$. Later on this sentence
will be made of rigorous meaning.
\medskip

\noindent$\bullet$\,\,\,Summarizing the above accepted
definitions, one easily represents `almost the whole' of the hydra
in the form
 \begin{equation*}
H^T_\gamma\setminus{\rm Corn
H^T_\gamma}\,=\,\bigcup\limits_{j=1}^J\bigcup\limits_{x\in\,
\omega\subset\, \Phi^j}{\mathscr
L}[\pi^{-1}(x)]=\bigcup\limits_{j=1}^J\bigcup\limits_{t\in\,
\psi\subset\, \Psi^j}{\mathscr L}[\rho^{-1}(t)]\,,
 \end{equation*}
which holds for any cells $\omega\subset\Phi^j$ and
$\psi\subset\Psi^j$.
\smallskip

\noindent$\bullet$\,\,\,The reason to introduce the partition
$\Pi$ and split the graph into families is that the waves $u^f$
depend on controls $f$ {\it locally} in the following sense. As
one can see from (\ref{Eq repres u single gamma}), the values
$u^f(\cdot,T)\big|_{\Phi}$ are determined by the values
$f\big|_{\Xi[\Phi]}$. Moreover,
 \begin{equation*}
{\rm supp\,}f\subset \Xi[\Phi]\quad\text{is equivalent to}\quad
{\rm supp\,}u^f(\cdot,T)\subset \Phi\,.
 \end{equation*}
Such a locality is helpful for analysis of the reachable sets. In
particular, it implies
 \begin{equation}\label{Eq locality reachable sets}
{\mathscr U}^T_\gamma\,=\,\oplus \sum_{\Phi\subset\Pi} {\mathscr
U}^T_\gamma\langle\Phi\rangle\,,
 \end{equation}
where ${\mathscr U}^T_\gamma\langle\Phi\rangle$ is the set of
waves supported in $\Phi$. Orthogonality of the sum is just a
consequence of $\Phi^j\cap\Phi^k=\emptyset$ for $j\not=k$.

\subsubsection*{Amplitude vectors}
A construction, which we introduce here, enables one to detail the
rep\-re\-sen\-ta\-tions of waves (\ref{Eq repres u single gamma})
and (\ref{Eq repres u general}).
\smallskip

\noindent$\bullet$\,\,\,Fix a point $x\in \Pi$; recall that such a
point belongs to the metric neigh\-bor\-hood $\Omega^T[\gamma]$
and lies outside the set of critical points. Let
$\Lambda[x]=\{x_k\}_{k=1}^{m_\Phi}$ be its determination set,
$\Phi=\bigcup_{k=1}^{m_\Phi}\omega_k$ be the family which $x$
belongs to (see (\ref{Eq Phi= sum Lambda}) and (\ref{Eq Pi= sum
Phi=sum Phi sum omega})). So, $x_k\in\omega_k$, whereas $x$
coincides with one of the points $x_k$. Let
 \begin{equation*}
\Xi[x]\,=\,\{t_i\}_{i=1}^{n_\Phi}:\,\,\,0<t_1<\dots<t_{n_\Phi}<T\,.
 \end{equation*}
Note that the values $t_i$ vary as $x$ varies over a cell of
$\Phi$ but the number $n_\Phi$ does not depend on $x$. Also notice
the evident equality $\Xi[x]=\Xi[x_k]$ for all $x_k\in\Lambda[x]$.
\smallskip

\noindent$\bullet$\,\,\,Now for an arbitrary $x\in
{\overline{\Omega^T[\gamma]}}\setminus\Gamma$ we put
 $$
m[x]:=\#\Lambda[x]\quad\text{and}\quad n[x]:=\#\Xi[x].
 $$
Note that for $x\in\Phi$ on has $m[x]=m_\Phi$ and $n[x]=n_\Phi$.
Then on the determination set one defines $n[x]$ functions
$\alpha^i: \Lambda[x]\to\mathbb R$ by
 \begin{equation}\label{Eq def ampl vectors}
\alpha^i(x_k)\,:=\,\begin{cases}
            a(x_k,t_i) & \text{if}\,\,\,(x_k,t_i)\in H^T_\gamma\\
            0          & \text{if}\,\,\,(x_k,t_i)\notin H^T_\gamma
              \end{cases},\qquad k=1,\dots,m[x]
 \end{equation}
and call them the {\it amplitude vectors}.

Return to representation (\ref{Eq repres u single gamma}). In
terms of the amplitude vectors, (\ref{Eq repres u single gamma})
can be written in the form
 \begin{equation*}
u^f(x_k,T)\big|_{x_k\in\Lambda[x]}\,=\,\sum\limits_{i=1}^{n[x]}\varphi(T-t_i)\alpha^i(x_k)\quad\qquad
(f=\delta_\gamma\varphi)\,,
 \end{equation*}
which represents the wave not only at $x$ but on the whole
determination set $\Lambda[x]$. In particular, varying $x$ over a
cell $\omega\subset\Phi$, we represent the wave $u^f(\cdot,T)$ on
the whole family $\Phi$ (see (\ref{Eq Phi= sum Lambda})).
\smallskip

\noindent$\bullet$\,\,\, Let $l_2(\Lambda[x])$ be the space of
functions on $\Lambda[x]$ with the inner product
 $$
\langle
f,g\rangle\,=\,\sum\limits_{x'\in\Lambda[x]}f(x')\,g(x')=\sum\limits_{k=1}^{m[x]}
f(x_k)\,g(x_k)\,.
 $$
It contains the subspace
 \begin{equation*}
{\mathbb A}[x]\,:=\,{\rm
span}\{\alpha^1,\dots,\alpha^{n[x]}\}\,,\qquad {\rm
dim\,}{\mathscr A}[x]\leqslant n[x]
 \end{equation*}
generated by the amplitude vectors.

In the sequel one makes the use of the more convenient basis in
${\mathbb A}[x]$. We redesign the system of amplitude vectors
$\alpha^1,\dots,\alpha^{n[x]}$ by the Schmidt procedure:
 \begin{equation}\label{Eq Schmidt}
\beta^{i}:=
  \begin{cases}
\frac{\alpha^1}{\|\alpha^1\|} & {\rm if}\,\,i=1\,,\\
\frac{\alpha^{i}-\sum \limits_{j=1}^{i-1}\langle \alpha^i,\,
\beta^{j}\rangle\,\beta^{j}}{\|\alpha^{i}-\sum
\limits_{j=1}^{i-1}\langle \alpha^{i},\,
\beta^{j}\rangle\,\beta^{j}\|}&{\rm if}\,\,i\geqslant 2\,\,\,{\rm
and}\,\,\,\alpha^{i} \not \in {\rm span\,}\{\alpha^{1}, \dots,
\alpha^{i-1}\}\\ 0 & {\rm otherwise}
  \end{cases},
 \end{equation}
and get a system $\beta^{1}, \dots, \beta^{n[x]}$. Its nonzero
elements satisfy $\langle \beta^{i}, \beta^{j}
\rangle=\delta_{ij}$, and ${\rm span\,}\{\beta^{1}, \dots,
\beta^{n[x]}\} = {\mathbb A}[x]$ holds.
\smallskip

\noindent$\bullet$\,\,\, For any two points
$x,x'\in\omega\subset\Phi$ one has
 \begin{equation*}\label{Eq alpha beta Piese-wise const}
\alpha^i(x) = \alpha^i(x') = \alpha^i\big|_{\omega}\,,\qquad
\beta^i(x) = \beta^i(x') = \beta^i\big|_{\omega}\,.
 \end{equation*}
Hence, $\alpha^i(\cdot)$ and $\beta^i(\cdot)$ are the piece-wise
constant functions on the family ${\Phi}$, these functions taking
constant values on the cells.

\subsubsection*{Projection $P^T_\gamma$}
Let $P^T_\gamma$ be an (orthogonal) projection in ${\mathscr
H}=L_2(\Omega)$ onto the reachable subspace ${\mathscr
U}^T_\gamma$. Here we provide a constructive description of this
projection.
\smallskip

\noindent$\bullet$\,\,\, For a subset $B \subset \Omega$, by
$\chi_B$ we denote its {\it indicator} (a characteristic function)
and introduce the subspace
$${\mathscr H}\langle B \rangle\,:=\chi_B {\mathscr H}=\{\chi_B y\,|\,\, y
\in \mathscr H\}$$ of functions supported on $B$. In accordance
with (\ref{Eq Pi= sum Phi=sum Phi sum omega}), one has
 \begin{equation}\label{H[Omega^T]= sum H <Phi>}
{\mathscr H}\langle \Omega^T[\gamma] \rangle=\oplus \sum
\limits_{\Phi \in \Pi} {\mathscr H}\langle \Phi \rangle\,, \quad
{\mathscr U}^T_\gamma \overset{(\ref{Eq locality reachable
sets})}= \oplus \sum \limits_{\Phi \in \Pi} {\mathscr
U}^T_\gamma\langle\Phi\rangle\,,
 \end{equation}
where ${\mathscr U}^T_\gamma\langle\Phi\rangle \subset {\mathscr
H}\langle\Phi\rangle$ is the subspace of waves supported in
$\Phi$. Therefore,
 \begin{equation}\label{P via Q Phi}
P^T_\gamma\,=\,\oplus\sum \limits_{\Phi \in \Pi}Q_\Phi\,,
 \end{equation}
where $Q_\Phi$ projects in ${\mathscr H}\langle \Phi \rangle$ onto
${\mathscr U}^T_\gamma\langle\Phi\rangle$. Hence, to characterize
$P^T_\gamma$ is to describe projections $Q_\Phi$.
\smallskip

\noindent$\bullet$\,\,\,As is shown in \cite{BWada 2}, the
projections $Q_{\Phi}$ are represented via the above introduced
vectors $\beta^i$ as follows:
 \begin{equation}\label{Q Phi via beta}
\left(Q_\Phi y\right)(x)\,=\,
  \begin{cases}
\sum \limits_{i=1}^{n_\Phi} \langle y\big|_{\Lambda[x]},
\beta^{i}\rangle\,\beta^{i}(x)\,, & x \in \Phi\\
0\,, & x \in \Omega \backslash \Phi
  \end{cases}\,,
 \end{equation}
where $y \in \mathscr H$ is arbitrary. So, recalling (\ref{P via Q
Phi}), we conclude that $P^T_\gamma$ is cha\-rac\-te\-ri\-zed.

\subsubsection*{Eikonal $E^T_\gamma$}
As well as the projection $P^T$, the eikonal $E^T_\gamma$ is
reduced by the subspaces ${\mathscr H}\langle\Phi\rangle$, i.e.,
$E^T_\gamma{\mathscr H}\langle\Phi\rangle\subset{\mathscr
H}\langle\Phi\rangle$ holds and implies
 \begin{equation}\label{Eq Eikonal reduced}
E^T_\gamma\,=\,\oplus\sum_{\Phi\subset\Pi}E^T_\gamma\langle\Phi\rangle
 \end{equation}
where $E^T_\gamma\langle\Phi\rangle:=E^T_\gamma\big|_{\Phi}$ is
the part of $E^T_\gamma$ acting in ${\mathscr
H}\langle\Phi\rangle$. As is shown in \cite{BWada 2}, the
representation
 \begin{equation}\label{Eq Eikonal repres H}
\left(E^T_\gamma\langle\Phi\rangle y\right)(x)\,=\, \begin{cases}
\sum \limits_{i=1}^{n_\Phi} \tau^i(x)\langle y\big|_{\Lambda[x]},
\beta^{i}\rangle\,\beta^{i}(x)\,, & x \in \Phi\\
0\,, & x \in \Omega \backslash \Phi
  \end{cases}
 \end{equation}
holds, where $y \in \mathscr H$ is arbitrary and $\tau^i$ are
defined by (\ref{Eq tau i}).

It is seen from (\ref{Eq Eikonal repres H}) that the eikonal is
also reduced by the parts of the reachable set (the summands in
(\ref{H[Omega^T]= sum H <Phi>})): one has
 \begin{equation*}
E^T_\gamma{\mathscr
U}^T_\gamma\langle\Phi\rangle\,\subset\,{\mathscr
U}^T_\gamma\langle\Phi\rangle\,,\,\, \qquad \Phi\subset\Pi\,.
 \end{equation*}

\subsubsection*{More hydras for $\gamma$}
For what follows it is reasonable to substitute $H^T_\gamma$ for
more convenient object which we call an efficient hydra and denote
by $\dot H^T_\gamma$. The latter is constructed via an auxiliary
extended hydra $\tilde H^T_\gamma$.
\smallskip

\begin{figure}
\centering \epsfysize=8cm \epsfbox{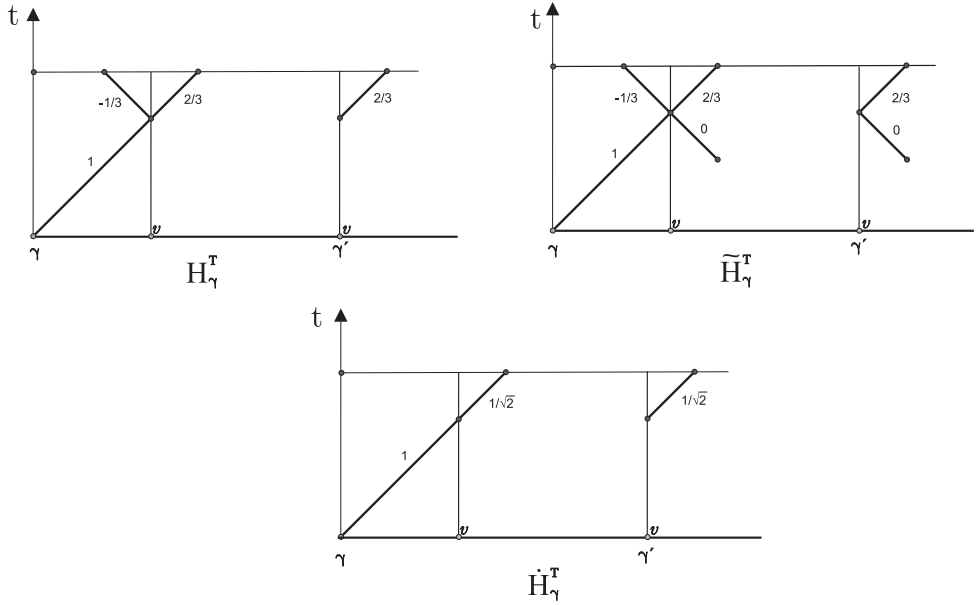} \caption[Fig.3]{The
original, extended, and efficient hydras}\label{The hydras}
 \end{figure}

\noindent$\bullet$\,\,\,Let $x\in\Omega^T[\gamma]$. Return to the
definitions (\ref{Eq def Lambda}), (\ref{Eq def Ksi}) and
introduce a {\it grid}
 $$
{\mathscr G}[x]\,:=\,\Lambda[x]\times \Xi[x]\,\supseteq {\mathscr
L}[\pi^{-1}(x)]\,.
 $$
Then define an {\it extended hydra}
 \begin{equation*}
\tilde
H^T_\gamma\,:=\,\overline{\bigcup\limits_{j=1}^J\bigcup\limits_{x\in\,
\omega\subset\, \Phi^j}{\mathscr G}[x]}\,\supseteq\,H^T_\gamma\,
 \end{equation*}
(see Fig. \ref{The hydras}). As is easy to recognize, the
neighborhood $\overset{\gamma}\sim$ and the lattices ${\mathscr
L}[B]$ are also well defined on $\tilde H^T_\gamma$. Along with
them, one defines the analogs of the sets (\ref{Eq def Lambda}),
(\ref{Eq def Ksi}), which obviously coincide with the original
$\Lambda[x]$,  $\Xi[x]$ and are denoted by the same symbols.

Extend the amplitude $a(x,t)$ from $H^T_\gamma$ to $\tilde
H^T_\gamma$ by zero and denote the extension by $\tilde a(x,t)$.
The extended amplitude is a piece-wise constant function on
$\tilde H^{T}_\gamma$.
\smallskip

\noindent$\bullet$\,\,\,Fix an $x\in\overline{\Omega^T[\gamma]}$
and define the new amplitude vectors $\tilde\alpha^i:
\Lambda[x]\to\mathbb R$ by
 \begin{equation*}
\tilde\alpha^i(x_k)\,:=\,\tilde a(x_k,t_i) \quad
\text{for}\,\,\,(x_k,t_i)\in \tilde H^{T}_\gamma, \qquad
x_k\in\Lambda[x]\,,
 \end{equation*}
where $t_i\in\Xi[x]=\Xi[x_k],\,i=1,\dots,n[x]$. Recall that
$n[x]=n_{\Phi}$ for $x\in\Phi$. Comparing with (\ref{Eq def ampl
vectors}), we see that these vectors coincide with the old ones
for $x\in\Phi$.

Applying the Schmidt process (\ref{Eq Schmidt}) to the system
$\{\tilde\alpha^i\}_{i=1}^{n[x]}$, we arrive at the vectors
$\beta^i: \Lambda[x]\to\mathbb R$. When $x$ varies over a cell
$\omega\subset\Phi$, the points $x_k$ also vary over the
corresponding cells $\omega_k\subset\Phi$ but the values of the
vector components $\beta^i(x_k)$ remain constant:
 \begin{equation}\label{Eq beta i
k=const} \beta^i(x_k)\,=:\,(\beta^i)_k\,={\rm const} \qquad
\text{as}\,\,x_k\in\omega_k\subset \Phi\,.
 \end{equation}

The vectors $\beta^i$ in turn determine a function $b$ on the
extended hydra by the following rule. Let $(x,t)\in\tilde
H^T_\gamma$ be such that $x=x_k\in\Lambda[x]$ and
$t=t_i\in\Xi[x]$; then we put $b(x,t)\,:=\,\beta^i(x_k)$. By its
definition, function $b$ is defined on the whole $\tilde
H^T_\gamma$. By its construction, $b$ is a piece-wise constant
function on the extended hydra.
\smallskip

\noindent$\bullet$\,\,\,Now we reduce $\tilde H^T_\gamma$ and turn
to an {\it efficient hydra}
 $$
\dot H^T_\gamma\,:=\,{\rm supp\,}b\,\subset \tilde H^T_\gamma\,.
 $$
The function $b\big|_{\dot H^T_\gamma}$ is said to be an efficient
amplitude.

For the pair $\{\dot H^T_\gamma, b\}$ one introduces the analogs
of all the objects, which were defined for $\{H^T_\gamma, a\}$.
Namely, one defines the space-time projections (\ref{Eq def pi rho
on H gamma}) for $\dot H^T_\gamma$ instead of $H^T_\gamma$, the
lattices ${\mathscr L}[B]\subset \dot H^T_\gamma$, the
neighborhood $\overset{\gamma}\sim$, determination sets (\ref{Eq
def Lambda}) and (\ref{Eq def Ksi}), corner and critical points
(\ref{Eq def Theta}), the set (\ref{Eq def Pi}), families (\ref{Eq
def Phi}) and their cells, and so on. In what follows we mark
these analogs by dot: $\dot\Lambda[x],\,\dot\Theta,
\,\dot\Pi,\,\dot\Phi,\,\dot\Psi,\dots$. The analogs are of the
same properties as the originals. In particular, the alternative
(\ref{Eq Lambda alternative}) holds for $\dot\Lambda[x]$.
\smallskip

\noindent$\bullet$\,\,\,The main reason to deal with $\dot
H^T_\gamma$ instead of $H^T_\gamma$ is that the partition
$\dot\Pi=\cup_j\,\Phi^j$ (with some new families $\Phi^j$ !)
provides the decompositions
 \begin{equation*}
{\mathscr H}\langle \Omega^T[\gamma] \rangle=\oplus \sum
\limits_{\Phi \in \dot\Pi} {\mathscr H}\langle \Phi \rangle\,,
\quad {\mathscr U}^T_\gamma = \oplus \sum \limits_{\Phi \in
\dot\Pi} {\mathscr U}^T_\gamma\langle\Phi\rangle\,,
 \end{equation*}
which are more natural and convenient for analysis of the eikonal
algebra. One more advantage is that the amplitude vectors of
$\{\dot H^T_\gamma,b\}$ defined by (\ref{Eq def ampl vectors})
(and denoted by $\dot\beta^i$ in the sequel), in essence, coincide
with the vectors $\beta^i$ (obtained by (\ref{Eq Schmidt})) and,
just by construction, constitute the orthogonal normalized bases
in the spaces $\dot{\mathscr A}[x]$ of functions on the
determination sets $\dot\Lambda[x]$.
\smallskip

\noindent$\bullet$\,\,\,As one can show, in terms of the efficient
hydra, the representations (\ref{P via Q Phi}) and (\ref{Q Phi via
beta}) take the form
 \begin{equation}\label{P via Q dot Phi}
P^T_\gamma\,=\,\oplus\sum \limits_{\Phi \in \dot\Pi}Q_\Phi\,,\quad
\left(Q_\Phi y\right)(x)\,=\,\begin{cases} \sum
\limits_{i=1}^{n_\Phi} \langle y\big|_{\dot\Lambda[x]},
\dot\beta^{i}\rangle\,\dot\beta^{i}(x)\,, & x \in \Phi\\
0\,, & x \in \Omega \backslash \Phi
  \end{cases}\,.
 \end{equation}
Respectively, for the eikonal, instead of (\ref{Eq Eikonal
reduced}) and (\ref{Eq Eikonal repres H}) we have
 \begin{align} \notag &
E^T_\gamma\,=\,\oplus\sum_{\Phi\subset\dot\Pi}E^T_\gamma\langle\Phi\rangle;\qquad
\left(E^T_\gamma\langle\Phi\rangle y\right)(x)\,=\\
\label{Eq Eikonal repres dot H} &=\, \begin{cases} \sum
\limits_{i=1}^{n_\Phi} \dot\tau^i(x)\langle
y\big|_{\dot\Lambda[x]},
\dot\beta^{i}\rangle\,\dot\beta^{i}(x)\,, & x \in \Phi\\
0\,, & x \in \Omega \backslash \Phi
  \end{cases}\,,
 \end{align}
where $\dot\tau^i$ are defined for $\dot H^T_\gamma$ in the same
way as $\tau^i$ for $H^T_\gamma$: see (\ref{Eq tau i}).

\subsubsection*{Partition $\Pi_\Sigma$}
Now, let $\Sigma=\{\gamma_1,\dots,\gamma_\sigma\}$ be a subset of
the graph boundary $\Gamma$. Recall that
$\Omega^T[\Sigma]\subset\Omega$ is a metric neighborhood of
$\Sigma$ of radius $T$. One has
$\Omega^T[\Sigma]=\cup_{\gamma\in\Sigma}\,\Omega^T[\gamma]$, so
that this neighborhood is a part of the graph filled (at the
moment $t=T$) by waves, which move from all $\gamma_k$.

In what follows we deal with a collection of hydras $\dot
H^T_\gamma$ for $\gamma\in\Sigma$. The objects related with single
vertices are marked by the subscript: $\dot\Lambda_\gamma[x],
\dot\Theta_\gamma, \dot\Pi_\gamma$ and so on.
\smallskip

\noindent$\bullet$\,\,\,Recall that (\ref{Eq def Phi}) and
(\ref{Eq Pi= sum Phi=sum Phi sum omega}) provide the partition of
$\Omega^T[\gamma]$ consistent with the structure of a single
hydra. Here we describe a partition of $\Omega^T[\Sigma]$ relevant
to the collection of the {\it efficient} hydras. Notice that it
differs from the one introduced in \cite{BWada 2} and associated
with the original hydras $H^T_\gamma$.

We say the points $x, x'\in\overline{\Omega^T[\Sigma]}$ to be the
space neighbors with respect to $\gamma$ and write
$x\overset{\gamma}\approx x'$ if
$\dot\Lambda_\gamma[x]=\dot\Lambda_\gamma[x']$. The relation
$\overset{\gamma}\approx$ is an equivalence.

The points $x, x'$ are said to be the space neighbors with respect
to $\Sigma$ (we write $x\overset{\Sigma}\approx x'$) if there are
the points $x_1,\dots, x_l$ and vertices $\gamma_{k_j}\in\Sigma$
such that $x\overset{\gamma_{k_1}}\approx
x_1\overset{\gamma_{k_2}}\approx
\dots\overset{\gamma_{k_l}}\approx
x_l\overset{\gamma_{k_{l+1}}}\approx x'$. The relation
$\overset{\Sigma}\approx$ is also an equivalence. By
$\Lambda_\Sigma[x]$ we denote the equivalence class of the point
$x$. Note the obvious relation
$\dot\Lambda_\gamma[x]\subset\Lambda_\Sigma[x]$ following from the
definitions.
\smallskip

\noindent$\bullet$\,\,\,Recall that
$\dot\Theta_\gamma\subset\overline{\Omega^T[\gamma]}$ is the set
of critical points determined by the hydra $\dot H^T_\gamma$, and
put $\Theta'_\Sigma:=\cup_{\gamma\in\Sigma}\dot\Theta_\gamma$.
Then we define
 \begin{equation*}
\Theta_\Sigma\,:=\,\bigcup\limits_{x\in\Theta'_\Sigma}\Lambda_\Sigma[x]\quad\text{and}\quad
\Pi_\Sigma\,:=\,\overline{\Omega^T_\Sigma}\setminus\Theta_\Sigma\,.
 \end{equation*}
It is the set $\Pi_\Sigma$ which provides the relevant partition
of the part $\Omega^T[\Sigma]$ filled by waves from $\Sigma$. As
is easy to verify, $\Pi_\Sigma$ consists of certain (new) families
$\Phi^j$, each family consisting of the cells of equal length:
 \begin{equation*}
\Pi_\Sigma\,=\,\bigcup\limits_{j=1}^{J_\Sigma}\Phi^j\,;\qquad\Phi^j=
\bigcup\limits_{i=1}^{m_j}\omega_i^j\,,\quad {\rm
diam\,}\omega^j_i=\epsilon_j\,.
 \end{equation*}
\smallskip

\noindent$\bullet$\,\,\,Now we modify the representations (\ref{P
via Q dot Phi}) and (\ref{Eq Eikonal repres dot H}) to make them
consistent with the partition $\Pi_\Sigma$.

Fix a point $x\in\Omega^T[\Sigma]$ and assume that
$x\in\Omega^T[\gamma]$. Let $x$ belong to a family
$\Phi\subset\dot\Pi_\gamma$ and let
$\dot\Lambda_\gamma[x]\subset\Phi$ be its determination set. Let
$\dot\beta^i_\gamma$ be the amplitude vectors on
$\dot\Lambda_\gamma[x]$. Recall the embedding
$\dot\Lambda_\gamma[x]\subset\Lambda_\Sigma[x]$ and extend all
$\dot\beta^i_\gamma$ from $\dot\Lambda_\gamma[x]$ to
$\Lambda_\Sigma[x]$ by zero. Simplifying the notation, we denote
these extensions by $\beta^i_\gamma$. Then one can get the
modified representation
 \begin{equation}\label{P via Q Phi Sigma}
P^T_\gamma\,=\,\sum \limits_{\Phi \in \Pi_\Sigma}Q_\Phi\,,\quad
\left(Q_\Phi y\right)(x)\,=\,
       \begin{cases}
   \sum
\limits_{i=1}^{n_\Phi} \langle y\big|_{\Lambda_\Sigma[x]},
\beta^{i}_\gamma\rangle\,\beta^{i}_\gamma(x)\,, & x \in \Phi\\
0\,, & x \in \Omega \backslash \Phi
       \end{cases}.
 \end{equation}
While (\ref{P via Q dot Phi}) and (\ref{Eq Eikonal repres dot H})
are related with $\dot\Pi_\gamma$, the new representation is
consis\-tent with the structure of $\Pi_\Sigma$ that occurs for
all $\gamma\in\Sigma$.
\smallskip

Respectively, one represents the eikonal as follows:
\begin{align}
\notag &
E^T_\gamma\,=\,\oplus\sum_{\Phi\subset\Pi_\Sigma}E^T_\gamma\langle\Phi\rangle;\qquad
\left(E^T_\gamma\langle\Phi\rangle y\right)(x)\,=\\
\label{Eq Eikonal repres H Sigma} &=\, \begin{cases} \sum
\limits_{i=1}^{n_\Phi} \tau^i_\gamma(x)\langle
y\big|_{\Lambda_\Sigma[x]},
\beta^{i}_\gamma\rangle\,\beta^{i}_\gamma(x)\,, & x \in \Phi\\
0\,, & x \in \Omega \backslash \Phi
  \end{cases}\,,
 \end{align}
where $y\in\mathscr H$ is arbitrary and $\tau^i_\gamma$ is
understood as $\dot\tau^i_\gamma$ extended from
$\dot\Lambda_\gamma[x]$ to $\Lambda_\Sigma[x]$ by zero.

\subsubsection*{Parametrization}
$\bullet$\,\,\,Choose a family
$\Phi=\cup_{k=1}^{m}\omega_k\subset\Pi_\Sigma$ \footnote{For a
wile, we simplify the notation:
$m=m_\Phi,\,\epsilon=\epsilon_\Phi,\,n=n_\Phi$ and so on.}; let
$\omega=]c,c'[\subset\Phi$ be one of the cells, which lies between
critical points $c$ and $c'$. Recall that all the cells are of the
same length $\epsilon=\tau(c,c')$, where $\tau$ is a distance on
the graph. For $x\in\omega$ we put $x=x(r)$ if $\tau(c,x)=r$.

Along with $x$, its determination set also turns out to be
parametrized: $\Lambda_\Sigma[x(r)]=\{x_k(r)\}_{k=1}^{m}$. As $r$
runs over $(0,\epsilon_\Phi)$, the points $x_k(r)$ vary
continu\-ous\-ly and sweep the cells $\omega_k$. Thus, the family
$\Phi$ is parametrized in whole.

By this, all elements of representations (\ref{P via Q Phi Sigma})
and (\ref{Eq Eikonal repres H Sigma}) are parametrized. The
vectors are
 $$
\beta^i_\gamma=\{(\beta^i_\gamma)_k(r)\}_{k=1}^{m},\,\,\,(\beta^i_\gamma)_k(r):=
\beta^i_\gamma(x_k(r))\overset{(\ref{Eq beta i
k=const})}=(\beta^i_\gamma)_k={\rm const},\quad
0<r<\epsilon_\Phi\,.
 $$
The functions take the form
$\tau^i_\gamma(r):=\tau^i_\gamma(x(r))$; as $r$ runs over
$(0,\epsilon_\Phi)$ the values $\tau^i_\gamma(r)$ sweep the proper
time cell $\psi_i=(t_{i-1},t_i)$. The definition (\ref{Eq tau i})
easily implies that
 \begin{equation}\label{Eq tau i gamma parametrized}
\text{either}\,\,\, \tau^i_\gamma(r)=t_{i-1}+r\quad \text{or}\quad
\tau^i_\gamma(r)=t_{i}-r=(t_{i-1}+\epsilon)-r
 \end{equation}
holds.

In the sequel we assume that each $\Phi\subset\Pi_\Sigma$ is
parametrized as described above.
\smallskip

\noindent$\bullet$\,\,\,With the parametrization one associates a
matrix representation of func\-ti\-ons and operators on the graph
as follows.

Let $\Phi\subset\Pi_\Sigma$ be a parametrized family,
$y\in\mathscr H$ a function on the graph, and let
$x=x(r)\in\Lambda_\Sigma[x(r)]=\{x_k(r)\}_{k=1}^{m}\subset\Phi,
\,0<r<\epsilon$. Introduce an isometry $U_\Phi$ by
\begin{equation*}
{\cal H}\langle \Phi \rangle \ni y
\,\,\,\overset{U_\Phi}\mapsto\,\,\,
  \begin{pmatrix}
y(x_1(r))\\\dots\\y(x_{m}(r))
  \end{pmatrix}\bigg|_{r \in (0, \,\epsilon)} \in L_2\left((0, \epsilon); {\mathbb
R}^{m}\right)\,.
 \end{equation*}
Define
\begin{equation*}
B_{\gamma}:=
   \begin{pmatrix}
(\beta^{1}_{\gamma})_1 & \dots & (\beta^{1}_{\gamma})_{m} \\
(\beta^{2}_{\gamma})_1 & \dots & (\beta^{2}_{\gamma})_{m} \\
\dots & \dots & \dots \\
(\beta^n_{\gamma})_1 & \dots & (\beta^n_{\gamma})_{m} \\
   \end{pmatrix},\,\,
D_{\gamma}(r):=\{\tau^i_\gamma(r)\,\delta_{ij}\}_{i,j=1}^{n}\,,\quad
r\in(0,\epsilon),
 \end{equation*}
where $\tau^i_\gamma(r)$  are of the form (\ref{Eq tau i gamma
parametrized}). Note that $B_{\gamma}^*B_{\gamma}$ is the constant
matrix which projects vectors in ${\mathbb R}^{m}$ onto the
subspace ${\mathbb A}_{\gamma}[x(r)]={\rm
span\,}\{{\beta}^1_\gamma, \dots , {\beta}^n_\gamma\}$. Along with
its generating vectors ${\beta}^i_\gamma$, this subspace does not
vary as $r$ runs over $(0,\epsilon)$. Therefore it is reasonable
to denote it by ${\mathbb A}_{\gamma}\langle\Phi\rangle$, what we
do in the sequel. So, we have
 $$
{\mathbb A}_{\gamma}\langle\Phi\rangle\,=\,{\rm
span\,}\{{\beta}^1_\gamma, \dots ,
{\beta}^n_\gamma\}\,=\,\left[B^*_\gamma B_\gamma\right] {\mathbb
R}^{m}\,.
 $$

\noindent$\bullet$\,\,\, The summand in (\ref{Eq Eikonal repres H
Sigma}) corresponding to the family $\Phi$ is represented as
follows:
 \begin{equation}\label{Eikonal 3}
\left(U_\Phi\, E^{\,T}_\gamma\!\langle\Phi\rangle\,
y\right)(r)\,=\,[B_{\gamma}^* D_{\gamma}(r) B_{\gamma}]\,
(U_\Phi{y})(r)\,, \qquad r \in (0, \epsilon)\,.
 \end{equation}

From here on, we assume that each family $\Phi\subset\Pi_\Sigma$
is parametrized by the proper $r\in(0,\epsilon_\Phi)$, so that the
matrices entering in (\ref{Eikonal 3}) are at our disposal. We
denote them by
 \begin{equation*}
B_{\gamma\,\Phi}=\{(\beta^i_{\gamma\,\Phi})_k\}_{{i=1,\dots,n_\Phi}
\atop k=1,\dots,m_\Phi} \quad\text{and}\quad D_{\gamma\,\Phi}={\rm
diag}\{\tau^i_{\gamma\,\Phi}(\cdot)\}_{i=1}^{m_\Phi}\,.
 \end{equation*}

\subsubsection*{Summary}
Let us resume the previous considerations.
\smallskip

\noindent$\bullet$\,\,\,Let
$\Sigma\subset\Gamma,\,\,\,\#\Sigma=:\sigma$. The collection of
efficient hydras $\{\dot H^T_\gamma\,|\,\,\gamma\in\Sigma\}$
deter\-mi\-nes the set of critical points $\Theta_\Sigma$. These
points divide the subdomain $\Omega^T[\Sigma]$ in parts
(families):
 \begin{equation*}
\Omega^T[\Sigma]\setminus\Theta_\Sigma=\Pi_\Sigma=\bigcup_{j=1}^J\Phi^j\,.
 \end{equation*}
This division determines the decomposition of subspaces
 \begin{equation*}
{\mathscr H}\langle \Omega^T[\Sigma] \rangle=\oplus \sum
\limits_{\Phi \in \Pi_\Sigma} {\mathscr H}\langle \Phi \rangle\,,
\quad {\mathscr U}^T_\Sigma = \oplus \sum \limits_{\Phi \in
\Pi_\Sigma} {\mathscr U}^T_\Sigma\langle\Phi\rangle\,,
 \end{equation*}
which reduces all the eikonals simultaneously:
 \begin{equation*}
E^T_\gamma {\mathscr H}\langle \Phi \rangle\subset {\mathscr
H}\langle \Phi \rangle\,, \,\,\, E^T_\gamma\,{\mathscr
U}^T_\Sigma\langle\Phi\rangle\subset{\mathscr
U}^T_\Sigma\langle\Phi\rangle\,,\qquad \gamma\in\Sigma\,.
 \end{equation*}
\noindent$\bullet$\,\,\,Parametrizing, we represent the subspaces
as
 \begin{equation*}
U_\Phi {\mathscr H}\langle \Phi \rangle =
L_2\left([0,\epsilon_\Phi]);{\mathbb R}^{m_\Phi}\right),\quad
U_\Phi {\mathscr U}^T_\Sigma\langle \Phi \rangle =
L_2\left([0,\epsilon_\Phi];{\mathbb A}_\Sigma\langle \Phi
\rangle\right)\,,
 \end{equation*}
where ${\mathbb A}_\Sigma\langle \Phi \rangle:={\rm
span}\{{\mathbb A}_\gamma\langle \Phi
\rangle\,|\,\,\gamma\in\Sigma\}$.

In the parametric form, the parts $E^T_\gamma\langle{\Phi}\rangle$
of the eikonals multiply elements of
$L_2((0,\epsilon_\Phi);{\mathbb R}^{m_\Phi})$ by the
matrix-functions (\ref{Eikonal 3}).

The total representation is realized by the operator
$U:=\oplus\sum_{\Phi\in\Pi_\Sigma}U_\Phi$ which provides
 \begin{align}
\notag & U{\mathscr H}\langle \Omega^T[\Sigma]
\rangle=\oplus\sum_{\Phi\in\Pi_\Sigma}L_2((0,\epsilon_\Phi);{\mathbb
R}^{m_\Phi})\,; \qquad
UE^T_\gamma U^{-1}\,=\\
\label{Eq Total repres Phi}
&=\,\oplus\sum_{\Phi\in\Pi_\Sigma}U_\Phi\,E^T_\gamma\!\langle\Phi\rangle\,
U^{-1}_\Phi\,=\,\oplus\sum_{\Phi\in\Pi_\Sigma}[B_{\gamma\,\Phi}^*
D_{\,\Phi}(\cdot) B_{\gamma\,\Phi}]\,, \qquad \gamma\in\Sigma\,.
 \end{align}
\noindent$\bullet$\,\,\,Turning to the eikonal algebra (\ref{Eq
def alg eik}), we have
 \begin{align}
\notag & {{\mathfrak
E}}^T_\Sigma\,=\,\vee\{E^T_\gamma\,|\,\,\gamma\in\Sigma\}\overset{(\ref{Eq
Eikonal repres H
Sigma})}=\,\vee\left\{\oplus\sum_{\Phi\subset\Pi_\Sigma}E^T_\gamma\langle\Phi\rangle\,\,\bigg|\,\,\gamma\in\Sigma\right\}\,,\\
\notag & U\,{\mathfrak E}^T_\Sigma U^{-1}\,\overset{(\ref{Eq Total
repres
Phi})}=\,\vee\left\{\oplus\sum_{\Phi\in\Pi_\Sigma}\left[B_{\gamma\,\Phi}^*
D_{\gamma\,\Phi}(\cdot)
B_{\gamma\,\Phi}\right]\,\bigg|\,\,\gamma\in\Sigma\right\}\,=\\
\label{Eq Total repres E T Sigma} & \overset{(\ref{Eq Eikonal
repres H
Sigma})}=\,\vee\left\{\oplus\sum_{\Phi\in\Pi_\Sigma}\left[\sum\limits_{i=1}^{n_\Phi}\tau^i_{\gamma\,\Phi}(\cdot)\,
P^i_{\gamma\,\Phi}\right]\,\bigg|\,\,\gamma\in\Sigma\right\} \,,
 \end{align}
where
$P^i_{\gamma\,\Phi}=\langle\cdot,\beta^i_{\gamma\,\Phi}\rangle\beta^i_{\gamma\,\Phi}$
are the constant (w.r.t. the parameter $r\in[0,\epsilon_\Phi]$)
one-dimensional matrix projections, the projections being
orthogonal by pairs:
$P^i_{\gamma\,\Phi}P^{i'}_{\gamma\,\Phi}=P^{i'}_{\gamma\,\Phi}P^i_{\gamma\,\Phi}=\mathbb
O$ for $i\not=i'$. The sum
$\oplus\sum_{i=1}^{n_\Phi}P^i_{\gamma\,\Phi}$ projects in
${\mathbb R}^{m_\Phi}$ onto the subspace  ${\mathbb
A}_{\gamma\,\Phi}={\rm
span}\{\beta^i_{\gamma\,\Phi}\,|\,\,i=1,\dots,n_\Phi\}$. In more
demonstrable form one has
 \begin{align}
\notag & U\,{\mathfrak E}^T_\Sigma U^{-1}= \vee\left\{\left(
\begin{array}{ccc}
\sum\limits_{i=1}^{n_{\Phi^1}} \tau^i_{\gamma\,\Phi^1}(\cdot_1) P^i_{\gamma\,\Phi^1}\\
& \ddots\\
&&\sum\limits_{i=1}^{n_{\Phi^J}} \tau^i_{\gamma\,\Phi^J}(\cdot_J)
P^i_{\gamma\,\Phi^J}
\end{array}
\right)\,\,\bigg|\,\,\,\,\gamma\in\Sigma\right\}\,\subset\\
\label{Eq demonstrable}& \subset \left(
\begin{array}{ccc}
C\left([0,\epsilon_1]; {\mathbb M}^{\,m_{\Phi^1}}\right)\\
& \ddots\\ &&C\left([0,\epsilon_J]; {\mathbb
M}^{\,m_{\Phi^J}}\right)
\end{array}
\right)
 \end{align}
where the arguments $\,\cdot_j\,$ run over $[0,\epsilon_j]$. Thus,
$U\,{\mathfrak E}^T_\Sigma U^{-1}$ is an operator algebra; its
elements multiply the elements of the representation space
 $$
{\mathscr S}^T_\Sigma\,:=\,\oplus\sum_{j=1}^J
L_2\left([0,\epsilon_{j}];{\mathbb R}^{m_{\Phi^j}}\right)
 $$
by the continuous matrix-valued functions of the proper structure.

\section{Algebra ${\mathfrak E}^T_{\Sigma}$}\label{sec Agebra Eik}
Representation (\ref{Eq Total repres E T Sigma}) enables one to
analyze the structure of the eikonal algebra, which is the main
subject of the paper. Analysis is preceded by some general facts
and results on algebras.

\subsubsection*{About C*-algebras}
Recall some of the definitions. We write ${\mathfrak
A}\cong{\mathfrak B}$ if the algebras ${\mathfrak A}$ and
${\mathfrak B}$ are isometrically isomorphic.
\smallskip

\noindent$\bullet$\,\,\,A C*-algebra is a Banach algebra with an
involution $x\mapsto x^*$ obeying
$(x^*)^*=x,\,\,(x+y)^*=x^*+y^*,\,\, (\lambda
x)^*={\lambda}x^*,\,\,(xy)^*=y^*x^*$, and
$\|x^*\|=\|x\|,\,\,\|x^*x\|=\|x\|^2$ holds\,\,\, \cite{Dix, Mur}.

A C*-algebra ${\mathfrak M}$ is {\it elementary} if there is a
Hilbert space ${\mathscr R}$ such that ${\mathfrak M}\cong
{\mathfrak S}_\infty({\mathscr R})$ holds, where ${\mathfrak
S}_\infty({\mathscr R})$ is the compact operator algebra in
${\mathscr R}$\,\,\cite{Dix}. We'll deal with ${\mathscr
R}={\mathbb R}^m$ and the matrix algebras ${\mathfrak M}$.
\smallskip

Let ${\mathscr T}$ be a topological space, $\{{\mathfrak
A}(t)\}_{t\in{\mathscr T}}$ a family of C*-algebras. The elements
$x\in\prod_{t\in{\mathscr T}}{\mathfrak A}(t)$, i.e., the
functions on ${\mathscr T}$ provided $x(t)\in{\mathscr T}$, are
called the {\it vector fields}.

A {\it continuous field of algebras} is a family $\{{\mathfrak
A}(t)\}_{t\in{\mathscr T}}$ endowed with a set of vector fields
$F\subset\{{\mathfrak A}(t)\}_{t\in{\mathscr T}}$ such that

\noindent$1.$\,\,\,$F$ is a linear space in $\{{\mathfrak
A}(t)\}_{t\in{\mathscr T}}$

\noindent$2.$\,\,\,for any $t\in{\mathscr T}$, the set
$\{x(t)\}_{x\in F}$ is dense in ${\mathfrak A}(t)$

\noindent$3.$\,\,\,for any $x\in F$, a function $t\mapsto\|x(t)\|$
is continuous in ${\mathscr T}$

\noindent$4.$\,\,\,if an element $x\in\prod_{t\in{\mathscr
T}}{\mathfrak A}(t)$ is such that for all
$\varepsilon>0,\,t\in{\mathscr T}$ there is a $\phi\in F$
providing $\|x(t)-\phi(t)\|<\varepsilon$, then $x\in F$.
\smallskip

\noindent$\bullet$\,\,\,The following fact plays the key role (see
\cite{Dix},\,\,sec 10.5.3).
\begin{Theorem}\label{Th 1}
Let ${\mathscr T}$ be a locally compact space, $({\mathfrak
A}(t),F)$ a continuous field of elementary C*-algebras on
${\mathscr T}$, ${\mathfrak A}$ the C*-algebra determined by this
field. Let ${\mathfrak B}\subset{\mathfrak A}$ be a C*-algebra
such that for any $t,t'\in{\mathscr T}$ and arbitrary
$\alpha\in{\mathfrak A}(t),\,\alpha'\in{\mathfrak A}(t')$ there is
an element $f\in{\mathfrak B}$ such that
$f(t)=\alpha,\,f(t')=\alpha'$ holds. Then ${\mathfrak
B}={\mathfrak A}$.
\end{Theorem}
We say that algebra $\mathfrak B$, possessing such a property,
{\it strongly separates} the points of ${\mathscr T}$.

In our case the fields $f$ will be the matrix-valued functions
given on a finite segment ${\mathscr T}=[0,\epsilon]$.

\subsubsection*{Standard algebras}
By ${\mathfrak B}({\mathscr G})$ we denote an algebra of bounded
operators acting in a Hilbert space ${\mathscr G}$. Let ${\mathbb
M}^n={\frak B}({\mathbb R}^n)$ be the algebra of $n\!\times\! n$ -
matrices with the norm $\|M\|={\rm sup}\{\|M\xi\|_{{\mathbb
R}^n}\,|\,\,\,\|\xi\|_{{\mathbb R}^n}=1\}$ and involution
(conjugation) $M\mapsto M^\ast$;
\smallskip

\noindent$\bullet$\,\,\,In the sequel we make use of the following
concrete algebras, which we call {\it standard}:

algebra $C[0,\epsilon]\equiv C\left([0,\epsilon]; {\mathbb
M}^1\right)$ of continuous functions with the norm
$\|f\|=\underset{0\leqslant t\leqslant\epsilon}{\rm sup\,}|f(t)|$
and its subalgebra $C_0[0,\epsilon]:=\{f\in
C[0,\epsilon]\,|\,\,f(0)=0\}$;
\smallskip

algebra $C\left([0,\epsilon]; {\mathbb M}^n\right)$ of continuous
matrix-valued functions with the norm $\|f\|=\underset{0\leqslant
t\leqslant\epsilon}{\rm sup}\|f(t)\|$, the point-wise (matrix)
multiplication and involution $f(t)\mapsto [f(t)]^\ast$;
\smallskip

(sub)algebra $\dot C\left([0,\epsilon]; {\mathbb
M}^3\right):=\left\{f \in C\left([0,\epsilon]; {\mathbb
M}^3\right)\,\big|\,\,\, f(0)\in {\mathbb M}^1\oplus{\mathbb
M}^2\right\}$;
\smallskip

operator algebra $C^{\rm op}\left([0,\epsilon]; {\mathbb
M}^n\right)\subset{\mathfrak B}\left(L_2([0,\epsilon];{\mathbb
R}^n)\right)\,\,\,(n\geqslant 1)$: its elements $f^{\rm op}$
multi\-ply vector-functions $u\in L_2([0,\epsilon];{\mathbb R}^n)$
by $f\in C\left([0,\epsilon]; {\mathbb M}^n\right)$;

operator (sub)algebra ${\dot C}^{\rm op}\left([0,\epsilon];
{\mathbb M}^n\right)$; its elements $f^{\rm op}$ multi\-ply
vector-functions $u\in L_2([0,\epsilon];{\mathbb R}^n)$ by $f\in
\dot C\left([0,\epsilon]; {\mathbb M}^n\right)$.
\smallskip

\noindent As is well known, $f\mapsto f^{\rm op}$ is an isometric
isomorphism of C*-algebras: $C\left([0,\epsilon]; {\mathbb
M}^n\right)\cong C^{\rm op}\left([0,\epsilon]; {\mathbb
M}^n\right) $ holds. By isometry, in what follows we identify
$C(\dots)\equiv C^{\rm op}(\dots)$.
\smallskip

\noindent$\bullet$\,\,\,More generally, we say a C*-algebra
$\mathfrak A$ to be standard if
 $$
{\mathfrak A}\,\cong\,\left\{f\in C\left([\alpha,\beta]; {\mathbb
P}\right)\,\big|\,\,\,f(\alpha)\in {\mathbb
P}_\alpha,\,\,f(\beta)\in{\mathbb P}_\beta\right\}\,,
 $$
where ${\mathbb P}_\alpha$ and ${\mathbb P}_\beta$ are the
C*-subalgebras of a matrix C*-algebra ${\mathbb P}\subseteq
{\mathbb M}^n$. General properties of such algebras are well
known: see \cite{Arv, Dix, Mur}. In what follows they play the
role of the `building blocks' which the eikonal algebra consists
of.

\subsubsection*{Block structure}
$\bullet$\,\,\,Return to (\ref{Eq Total repres E T Sigma}),
(\ref{Eq demonstrable}) and define
\begin{align}\label{Eq orthogonality}
E_{\gamma\,\Phi}(r):=\sum\limits_{i=1}^{n_\Phi}\tau^i_{\gamma\,\Phi}(r)
P^i_{\gamma\,\Phi};\,\,\,{\mathfrak
b}_\Phi(r):=\,\vee\{E_{\gamma\,\Phi}(r)\,|\,\,\gamma\in\Sigma\},\,\,\,
r\in[0,\epsilon_\Phi]\,.
 \end{align}

Looking at the form of (\ref{Eq Total repres E T Sigma}) and
(\ref{Eq demonstrable}), it is reasonable to say the algebra
 $
{\mathfrak b}_\Phi(\cdot)
 $
to be a {\it block}\, of the algebra $U\,{\mathfrak E}^T_\Sigma
U^{-1}$ corresponding to the family $\Phi$. Deno\-ting ${\mathbb
P}_\Phi:=\vee\{P^i_{\gamma\,\Phi}\,|\,\,i=1,\dots,n_\Phi;\,\,\gamma\in\Sigma\}\subset{\mathbb
M}^{\,m_\Phi}$, we have the evident relations
 $$
{\mathfrak
b}_\Phi(\cdot)\,\subset\,C\left([0,\epsilon_\Phi];{\mathbb
P}_\Phi\right)\,\subset\,C\left([0,\epsilon_\Phi];{\mathbb
M}^{\,m_\Phi}\right)\,.
 $$
By ${\mathfrak b}_\Phi\big|_K$ we denote the set of restrictions
of matrix-functions $b\in{\mathfrak b}_\Phi$ onto the subset
$K\subset[0,\epsilon_\Phi]$.
 \begin{Lemma}\label{Lemma 1}
For any $[a,b]\subset(0,\epsilon_\Phi)$ the relation ${\mathfrak
b}_\Phi\big|_{[a,b]}\,=\,C\left([a,b];{\mathbb P}_\Phi\right)$ is
valid.
 \end{Lemma}
$\square$\,\,\,For a function $\tau=\tau(s)$, by $\tau(K)\subset
\mathbb R$ we denote the range of its values as $s$ varies over
$K$.
\smallskip

\noindent$\bf \ast$\,\,\,Fix $\Phi,\,\gamma$ and $i$ for a while.
Take the values $r',\,r''$ of the parameter $r$ provided
$0<a\leqslant r'<r''\leqslant b<\epsilon_\Phi$. The property
(\ref{Eq tau i gamma parametrized}) implies
 $$
\tau^i_{\gamma\,\Phi}([a,b])\cap\tau^{l}_{\gamma\,\Phi}([a,b])=\emptyset
\qquad \text{for}\,\,\,l\not=i\,.
 $$
Hence, there is a polynomial
$q=q(\lambda)\,\,\,(\lambda\in{\mathbb R})$ such that
 \begin{align*}
& q(0)=0;\quad
q\big|_{\lambda=\tau^{l}_{\gamma\,\Phi}(r')}=0\,\,\,\text{for
all}\,\,\,l=1,\dots,n_\Phi; \quad
q\big|_{\lambda=\tau^{l}_{\gamma\,\Phi}(r'')}=0\,\,\,\text{for}\,\,l\not=i;\\
& q\big|_{\lambda=\tau^{i}_{\gamma\,\Phi}(r'')}=1\,.
 \end{align*}

\noindent$\bf \ast$\,\,\,By orthogonality of the projections
$P^i_{\gamma\,\Phi}$ in (\ref{Eq orthogonality}) we have
 $$
\left(q\left(E_{\gamma\,\Phi}\right)\right)(r)\,:=\,\sum\limits_{l=1}^{n_\Phi}q\left(\tau^l_{\gamma\,\Phi}(r)\right)
P^l_{\gamma\,\Phi}
 $$
that leads to
\begin{align}\label{Eq single proj}
\left(q\left(E_{\gamma\,\Phi}\right)\right)(r')=\mathbb
O\,\,\,\text{and}\,\,\,
\left(q\left(E_{\gamma\,\Phi}\right)\right)(r'')=P^i_{\gamma\,\Phi}
 \end{align}
owing to the choice of the polynomial.
\smallskip

\noindent$\bf \ast$\,\,\,By arbitrariness of $a,\,b,\,r''$ the
second relation in (\ref{Eq single proj}) means that each single
projection $P^i_{\gamma\,\Phi}$ belongs to the algebra ${\mathfrak
b}_\Phi(r)$ generated by `sums of such projections'. Therefore we
get
 \begin{align}\label{Eq b(r) into [a,b]}
& {\mathfrak
b}_\Phi(r)=\,\vee\{P^i_{\gamma\,\Phi}\,|\,\,i=1,\dots,n_\Phi;\,\,\gamma\in\Sigma\}={\mathbb
P}_\Phi,\quad r\in(0,\epsilon_\Phi)\,.
 \end{align}
The second consequence of (\ref{Eq single proj}) is the following.
It is easy to see that the proper choice of the polynomial $q$
enables one to change the roles of $r',\,r''$ and get the relation
$\left(q\left(E_{\gamma\,\Phi}\right)\right)(r'')=\mathbb
O\,\,\,\text{and}\,\,\,
\left(q\left(E_{\gamma\,\Phi}\right)\right)(r')=P^i_{\gamma\,\Phi}$.
Then, combining the relations, one can find a polynomial $\tilde
q$ provided $\tilde q(0)=0$, which obeys $\tilde q(r)=p$ and
$\tilde q(r')=p'$ for {\it any} $r,r'\in(0,r_\Phi)$ and
$p,p'\in{\mathbb P}_\Phi$. The latter means that the (sub)algebra
${\mathfrak b}_\Phi\big|_{[a,b]}\subset C\left([a,b];{\mathbb
P}_\Phi\right)$ strongly separates points of the compact
${\mathscr T}=[a,b]$. Applying Theorem \ref{Th 1}, we conclude
that the subalgebra exhausts the algebra, what the Lemma
claims.\qquad$\blacksquare$
\medskip

Notice in addition that (\ref{Eq b(r) into [a,b]}) may be invalid
at the endpoints $r=0, \epsilon_\Phi$. For instance, if
$\tau^i_{\gamma\,\Phi}(0)=0$ then the projection
$P^i_{\gamma\,\Phi}$ drops out of the generators of ${\mathfrak
b}_\Phi(0)$ that may lead to ${\mathfrak b}_\Phi(0)\not={\mathbb
P}_\Phi$. Similarly, if
$\tau^i_{\gamma\,\Phi}(\epsilon_\Phi)=\tau^{i+1}_{\gamma\,\Phi}(\epsilon_\Phi)$
then in (\ref{Eq orthogonality}) one gets
 $$
E_{\gamma\,\Phi}(\epsilon_\Phi)=\dots+\tau^i_{\gamma\,\Phi}(\epsilon_\Phi)
[P^i_{\gamma\,\Phi}+P^{i+1}_{\gamma\,\Phi}]+\dots
 $$
that also reduces the list of generators and may lead to
${\mathfrak b}_\Phi(\epsilon_\Phi)\not={\mathbb P}_\Phi$. One more
occasion is the equalities like
$\tau^i_{\gamma\,\Phi'}(\epsilon_{\Phi'})=\tau^i_{\gamma\,\Phi''}(0)$
which provide certain connections between the different families.
These effects do occur in the known examples and will be
demonstrated in section \ref{sec Simple graph}.
\smallskip

\noindent$\bullet$\,\,\,So, each block-algebra ${\mathfrak
b}_\Phi$ consists of the continuous ${\mathbb P}_\Phi$-valued
functions satisfying certain conditions at the endpoints of
$[0,\epsilon_\Phi]$. Respectively, the bulk of the algebra
$U{\mathfrak E}^T_\Sigma U^{-1}\cong{\mathfrak E}^T_\Sigma$ is
exhausted by the sum
$\oplus\sum\limits_{\Phi\subset\Pi_\Sigma}C\left((0,\epsilon_\Phi);
{\mathbb P}_\Phi \right)$. More subtle considerations are required
to clarify possible connections between the summands at the
endpoints $r=0,\epsilon_\Phi$. As a result, with some abuse of
terms, we can claim that the eikonal algebra consists of the {\it
standard algebras}.

\subsubsection*{On matrix algebras}
Dealing with concrete examples, we have to reveal the structure of
the algebras ${\mathbb P}_\Phi\subset {\mathbb M}^{\,m_\Phi}$. By
doing so, we regard ${\mathbb M}^{\,n}$ as an algebra of operators
acting in ${\mathbb R}^n$ and make use of the following simple
facts.
\smallskip

\noindent$\bf 1.$\,\,\,Any C*-subalgebra of the algebra
$\mathbb{M}^n$ is isometrically isomorphic to a sum
$\oplus\sum_{k}\mathbb{M}^{n_k}$, where $\sum{n_k}\leqslant n$.
\smallskip

\noindent$\bf 2.$\,\,\,The only irreducible C*-subalgebra of the
algebra $\mathbb{M}^n$ is $\mathbb{M}^n$ itself.
\smallskip

\noindent$\bf 3.$\,\,\,Let $P_1,\dots,P_\sigma \subset
\mathbb{M}^n$ be a set of projections: $P_l^*=P_l=P^2_l$. Then the
equality $\vee\{P_1,\dots,P_\sigma\}=\mathbb{M}^n$ is valid if and
only if 
no one nonzero vector in ${\mathbb R}^n$ is an eigenvector for all
$P_j$ simultaneously.
\smallskip

\noindent$\bf 4.$\,\,\,If $E_l=\sum_{i=1}^{n_l}\lambda_{l}^i
P^i_l\,\,\,\,(l=1,\dots,\sigma;\,\lambda_{l}^i\not=0)$ is the
spectral decomposition, then the equality
$\vee\{E_1,\dots,E_\sigma\}=\vee\{P^i_l\,|\,\,i=1,\dots,n_l;\,\,\,l=1,\dots,\sigma\}$
holds.

\section{Simple graph}\label{sec Simple graph}
As a simple (but not trivial!) example, we deal with a graph
$\Omega$, which consists of three edges $e_1,\,e_2,\,e_3$, three
boundary vertices вершин $\gamma_1,\,\gamma_2,\,\gamma_3,\,$ and a
single interior vertex $v$. The lengths of edges satisfy a generic
condition
 $$
l_1<l_2< l_3\,.
 $$
We take $\Sigma=\{\gamma_1,\gamma_2\}$ and study the structure of
${\mathfrak E}^T_{\Sigma}$ and evolution of this structure w.r.t.
$T$. Namely, we search the cases $T=T_{1,\,2,\,3,\,4}$ such that

\noindent$\bullet$\,\,\,\,\,$T_1<l_1$

\noindent$\bullet$\,\,\,\,\,$l_1<T_2<\frac{l_1+l_2}{2}$

\noindent$\bullet$\,\,\,\,\,$\frac{l_1+l_2}{2}<T_3<l_2$

\noindent$\bullet$\,\,\,\,\,$l_2<T_4<l_1+l_2$\,.
\bigskip

In the rest of the paper we omit technical details and just
present the results. All of them are simply verifiable.

On default, the absent matrix entries are assumed equal to zero.

\subsection*{The moment $T_1$}
At this moment the domains on the graph filled by waves are
disposed so that
$\Omega^{T_1}[\gamma_1]\cap\Omega^{T_1}[\gamma_2]=\emptyset$ and
$v$ is not captured by waves:
$v\not\in\Omega^{T_1}[\gamma_{1,2}]$. The partition of the filled
domain is $\Pi_\Sigma=\Phi^1\cup\Phi^2$: see Fig. \ref{The moment
$T_1$}, \ref{The moment $T_1$; the hydras}

The families are parametrized so that

\noindent\,\,\,$\tau^1_{\gamma_1\,\Phi^1}(r_1)=r_1\!:\quad
r_1\in[0,T_1]$

\noindent\,\,\,$\tau^1_{\gamma_2\,\Phi^2}(r_2)=r_2\!:\quad
r_2\in[0,T_1]$

\noindent holds. Note that here we have
$\tau^i_{\gamma_l\,\Phi^j}(0)=0$.
\smallskip

Respectively, the matrices (\ref{Eq orthogonality}) which
represent the eikonals, take the form
 $$
E_{\gamma_1\,\Phi^1}=\left(\begin{array}{cc}
r_1 & \\
 & 0
\end{array}\right), \quad E_{\gamma_2\,\Phi^2}=\left(\begin{array}{cc}
0 & \\
 & r_2
\end{array}\right)\,.
 $$
As a result, for the eikonal algebra we easily get
 $$
{\mathfrak
E}^{T_1}_{\{\gamma_1,\,\gamma_2\}}=\vee\{E^{T_1}_{\gamma_1},\,E^{T_1}_{\gamma_2}\}\cong\,
C_0[0,T_1]\oplus C_0[0,T_1].
 $$
 \begin{figure}
\centering \epsfysize=4cm \epsfbox{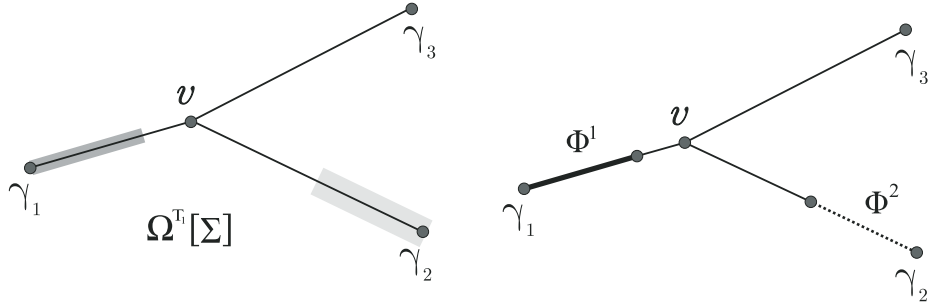}
\caption[Fig.4]{$T=T_1$; the graph}\label{The moment $T_1$}
 \end{figure}

 \begin{figure}
\centering \epsfysize=5cm \epsfbox{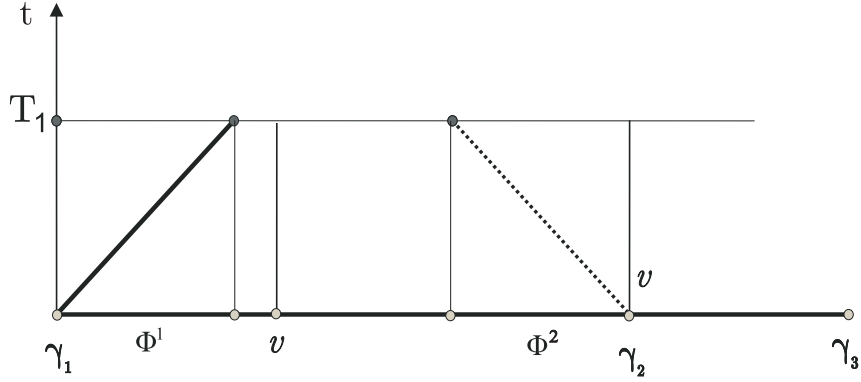}
\caption[Fig.5]{$T=T_1$; the hydras}\label{The moment $T_1$; the
hydras}
 \end{figure}

\subsection*{The moment $T_2$}
At this moment one has
$\Omega^{T_2}[\gamma_1]\cap\Omega^{T_2}[\gamma_2]=\emptyset$,
$v\in\Omega^{T_2}[\gamma_1]$ and $v\not\in\Omega^{T_2}[\gamma_2]$,
so that the waves from $\gamma_2$ do not reach the interior vertex
yet.
 \begin{figure}
\centering \epsfysize=4cm \epsfbox{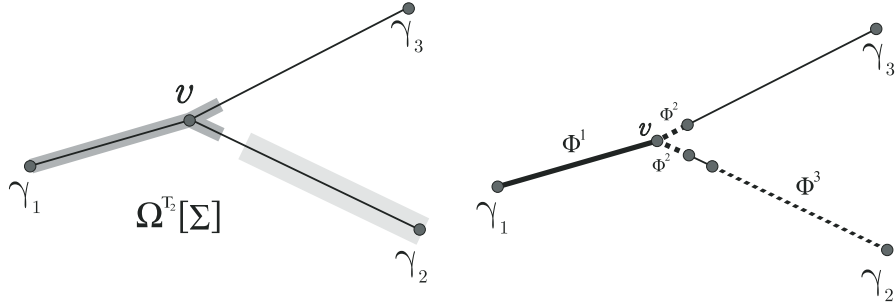}
\caption[Fig.6]{$T=T_2$; the graph}\label{The moment $T_2$}
 \end{figure}
In this case one has $\Pi_\Sigma=\Phi^1\cup\Phi^2\cup\Phi^3$: see
Fig. \ref{The moment $T_2$}, \ref{The moment $T_2$; the hydras}.
 \begin{figure}
 \centering \epsfysize=6cm \epsfbox{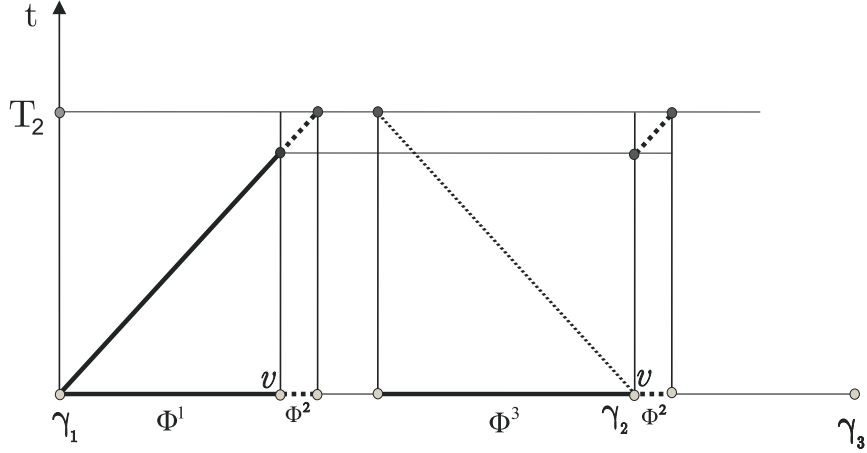}
\caption[Fig.7]{$T=T_2$; the hydras}\label{The moment $T_2$; the
hydras}
 \end{figure}

\noindent$\bullet$\,\,\,The proper parametrization provides

\noindent\,\,\,$\tau^1_{\gamma_1\,\Phi^1}(r_1)=r_1\!:\quad
r_1\in[0,l_1]$;

\noindent\,\,\,$\tau^1_{\gamma_1\,\Phi^2}(r_2)=l_1+r_2\!:\quad
r_2\in[0,T_2-l_1]$;

\noindent\,\,\,$\tau^1_{\gamma_2\,\Phi^3}(r_3)=r_3\!:\quad
r_3\in[0,T_2]$.
\smallskip

\noindent Notice the equalities
 \begin{equation}\label{Eq matching}
\tau^1_{\gamma_1\,\Phi^1}(0)=\tau^1_{\gamma_2\,\Phi^3}(0)=0;
\qquad
\tau^1_{\gamma_1\,\Phi^1}(l_1)=\tau^1_{\gamma_1\,\Phi^2}(0)\,,
 \end{equation}
which will impose certain matching conditions on the
matrix-functions.
\smallskip

\noindent$\bullet$\,\,\,The matrices (\ref{Eq orthogonality}) are
of the form
 $$
E_{\gamma_1\,\Phi^1}=r_1 P^1_{\gamma_1\,\Phi^1},\quad
E_{\gamma_1\,\Phi^2}=(l_1+r_2)P^1_{\gamma_1\,\Phi^2},\,\quad
E_{\gamma_2\,\Phi^3}= r_3 P^1_{\gamma_2\,\Phi^3}\,,
 $$
where
 \begin{align*}
P^1_{\gamma_1\,\Phi^1}=
 \left(\begin{array}{cccc}
1  \\
 & 0\\
&& 0\\
&&&0\\
\end{array}\right), \, P^1_{\gamma_1\,\Phi^2}=
\left(\begin{array}{cccc}
0  \\
 & \frac{1}{2} & \frac{1}{2}\\
& \frac{1}{2} & \frac{1}{2}\\
&&&0\\
\end{array}\right),\, P^1_{\gamma_2\,\Phi^3}=
 \left(\begin{array}{cccc}
0  \\
 & 0\\
&& 0\\
&&& 1\\
\end{array}\right).
 \end{align*}
Then one gets the representatives of the eikonals:
 \begin{align}
\notag & UE^{T_2}_{\gamma_1}U^{-1} =
E_{\gamma_1\,\Phi^1}+E_{\gamma_1\,\Phi^2}=\\
\label{Eq T2 1} & =r_1
P^1_{\gamma_1\,\Phi^1}+(l_1+r_2)P^1_{\gamma_1\,\Phi^2}=
\left(\begin{array}{cccc}
r_1  \\
 & \frac{l_1+r_2}{2} & \frac{l_1+r_2}{2}\\
& \frac{l_1+r_2}{2} & \frac{l_1+r_2}{2}\\
&&&0\\
\end{array}\right), \\
\label{Eq T2 2} &
UE^{T_2}_{\gamma_2}U^{-1}=E_{\gamma_2\,\Phi^3}=r_3
P^1_{\gamma_2\,\Phi^3}=
 \left(\begin{array}{cccc}
0 & \\
&0\\
&&0\\
 &&& r_3
\end{array}\right)\,.
 \end{align}
\noindent$\bullet$\,\,\,By orthogonality of the projections
$P^1_{\gamma\,\Phi}$, for any polynomial $q=q(\lambda)$ provided
$q(0)=0$ we have
 $$
Uq(E^{T_2}_{\gamma_1})U^{-1} =
q(r_1)P^1_{\gamma_1\,\Phi^1}+q(l_1+r_2)P^1_{\gamma_1\,\Phi^2}
 $$
that easily leads to
 \begin{align*}
&
U\left[\vee\{E^{T_2}_{\gamma_1}\}\right]U^{-1}\cong\\
& \cong\left\{\left(\begin{array}{cc}
\phi(r_1) & \\
 & \psi(r_2)
\end{array}\right)\bigg|\,\,\,\phi\in C_0[0,l_1],\,\psi\in
 C[0,T_2-l_1],\,\,\phi(l_1)\overset{(\ref{Eq
 matching})}=\psi(0)\right\}\cong\\
& \cong C_0[0,T_2]\,.
 \end{align*}
In the mean time one has
 $$
U\left[\vee\{E^{T_2}_{\gamma_2}\}\right]U^{-1}\cong
C_0[0,T_2]\,.
 $$
Summarizing, we represent the eikonal algebra in the form
 $$
{\mathfrak E}^{T_2}_{\{\gamma_1,\,\gamma_2\}}\cong
C_0[0,T_2]\oplus C_0[0,T_2]\,.
 $$

\subsection*{The moment $T_3$}
At this moment one has
$\Omega^{T_3}[\gamma_1]\cap\Omega^{T_3}[\gamma_2]\not=\emptyset$,
$v\in\Omega^{T_3}[\gamma_1]$, $v\not\in\Omega^{T_3}[\gamma_2]$.
The waves from $\gamma_2$ do not reach the interior vertex yet but
overlap with the waves going from $\gamma_1$. So, at $t=T_3$ the
waves from different boundary vertices begin to interact. As will
be seen, interaction leads to the curious effect: while
${\mathfrak E}^{T_1}_\Sigma$ and ${\mathfrak E}^{T_2}_\Sigma$ are
commutative, for $T\geqslant T_3$ the algebra ${\mathfrak
E}^{T}_\Sigma$  becomes {\it noncommutative}.
\smallskip

\begin{figure}[h!]
\centering \epsfysize=4cm \epsfbox{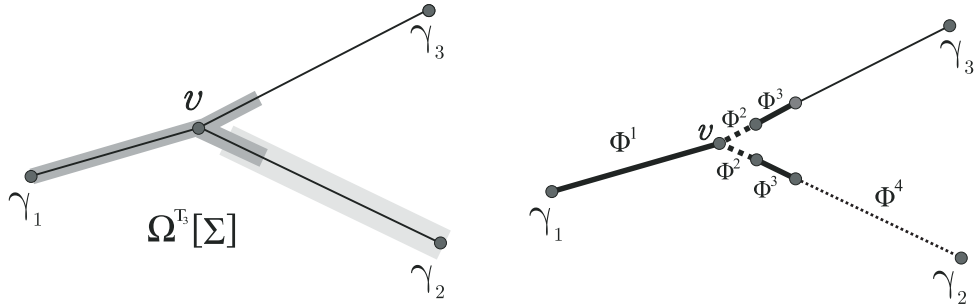}
\caption[Fig.8]{$T=T_3$; the graph}\label{$T=T_3$; the graph}
 \end{figure}

\begin{figure}[h!]
\centering \epsfysize=6cm \epsfbox{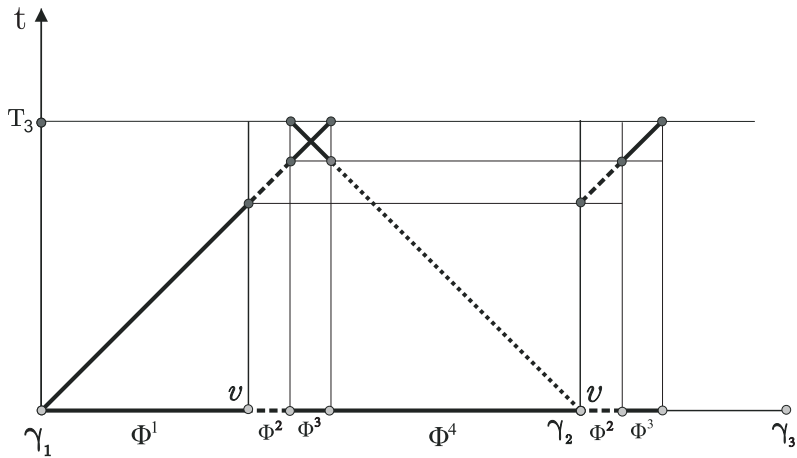}
\caption[Fig.9]{$T=T_3$; the hydras}\label{$T=T_3$; the hygras}
 \end{figure}

\noindent$\bullet$\,\,\,Now the partition of the domain filled by
waves is $\Pi_\Sigma=\Phi^1\cup\Phi^2\cup\Phi^3\cup\Phi^4$: see
Fig. \ref{$T=T_3$; the graph}, \ref{$T=T_3$; the hygras}.

The proper parametrization provides

\noindent\,\,\,$\tau^1_{\gamma_1\,\Phi^1}(r_1)=r_1\!:$\quad
$r_1\in[0,l_1]$;
\smallskip

\noindent\,\,\,$\tau^1_{\gamma_1\,\Phi^2}(r_2)=l_1+r_2\!:$\quad
$r_2\in[0,l_2-T_3]$;
\smallskip

\noindent\,\,\,$\tau^1_{\gamma_1\,\Phi^3}(r_3)=(l_1\!+\!l_2\!-\!T_3)+r_3,\,\,\,\tau^1_{\gamma_2\,\Phi^3}(r_3)=
T_3-r_3\!:$\quad $r_3\in[0,2T_3\!-\!l_1\!-\!l_2]$;
\smallskip

\noindent\,\,\,$\tau^1_{\gamma_2\,\Phi^4}(r_4)=r_4\!:$\quad
$r_4\in[0,l_1\!+\!l_2\!-\!T_3]$
\smallskip

\noindent$\bullet$\,\,\,As one can derive, the generators of the
algebra $UE^{T_3}_{\Sigma}U^{-1}$ take the form
 \begin{align}
\label{Eq T3 1}&
UE^{T_3}_{\gamma_1}U^{-1}=\left(\begin{array}{cccccc}
r_1  \\
 & \frac{l_1+r_2}{2} & \frac{l_1+r_2}{2}\\
 & \frac{l_1+r_2}{2} & \frac{l_1+r_2}{2}\\
 &&&\frac{(l_1+l_2-T_3)+r_3}{2} & \frac{(l_1+l_2-T_3)+r_3}{2}\\
 &&&\frac{(l_1+l_2-T_3)+r_3}{2} & \frac{(l_1+l_2-T_3)+r_3}{2}\\
 &&&&&0
\end{array}\right)\,,\\
\label{Eq T3 2}&
UE^{T_3}_{\gamma_2}U^{-1}=\left(\begin{array}{cccccc}
 0 & \\
 &0\\
 &&0\\
 &&&0&0\\
 &&&0&
T_3\!-\!r_3\\
 &&&&& r_4
\end{array}\right)\,.
 \end{align}
Looking at their structure, one can select the blocks
$\{(\cdot)_{i\,j}\}_{i,j=1}^3$ and $\{(\cdot)_{6\,6}\}$, which act
in perfect analogy to (\ref{Eq T2 1}) and (\ref{Eq T2 2}). These
blocks generate the `commutative part' of
$UE^{T_3}_{\Sigma}U^{-1}$, the part being isometrically isomorphic
to $C_0[0,l_1\!+\!l_2\!-\!T_3]\oplus C_0[0,l_1\!+\!l_2\!-\!T_3]$.

The origin of noncommutativity is the presence of the bloks
$\{(\cdot)_{i\,j}\}_{i,j=4}^5$ proportional to the projections
 \begin{align*}
 p=\left(\begin{array}{cc}
 \frac{1}{2} & \frac{1}{2}\\
 \frac{1}{2} & \frac{1}{2}\\
\end{array}\right)\quad \text{and} \quad p'=\left(\begin{array}{cc}
 0&0\\
 0&1
 \end{array}\right)\,,
 \end{align*}
which do not commute. These projections nave no mutual
eigenvectors. Therefore one has $\vee\{p,\,p'\}={\mathbb M}^2$ by
the property $3.$ of the matrix algebras (see the last item of
sec. \ref{sec Agebra Eik}).
\smallskip

\noindent$\bullet$\,\,\,As a result, one can arrive at the final
representation
 $$
{\mathfrak E}^{T_3}_{\{\gamma_1,\,\gamma_2\}}\cong
C_0[0,l_1\!+\!l_2\!-\!T_3]\oplus C_0[0,l_1\!+\!l_2\!-\!T_3]\oplus
С([0,2T_3\!-\!l_1\!-\!l_2],{\mathbb M}^2).
 $$
Notice that some work has to be done to check that the summands
are independent, i.e., there are no matching conditions at the
endpoints of the segments $[0,\epsilon_j]$ which might connect the
values of matrix-functions belonging to different summands.

\subsection*{The moment $T_4$}
\begin{figure}[h!]
\centering \epsfysize=4cm \epsfbox{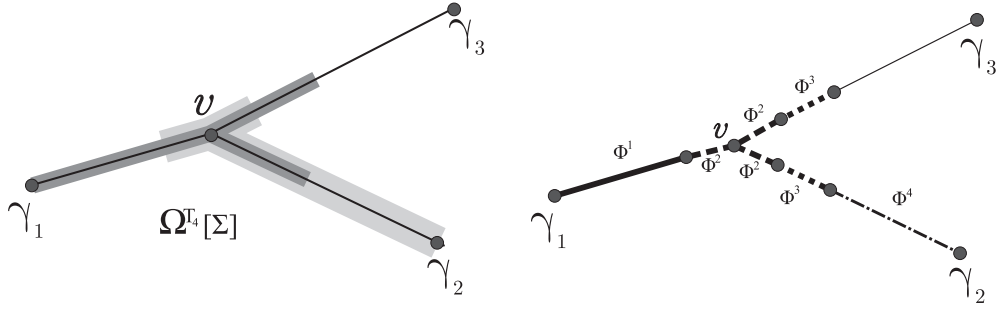}
\caption[Fig.10]{$T=T_4$; the graph}\label{$T=T_4$; the graph}
 \end{figure}

\begin{figure}[h!]
\centering \epsfysize=6cm \epsfbox{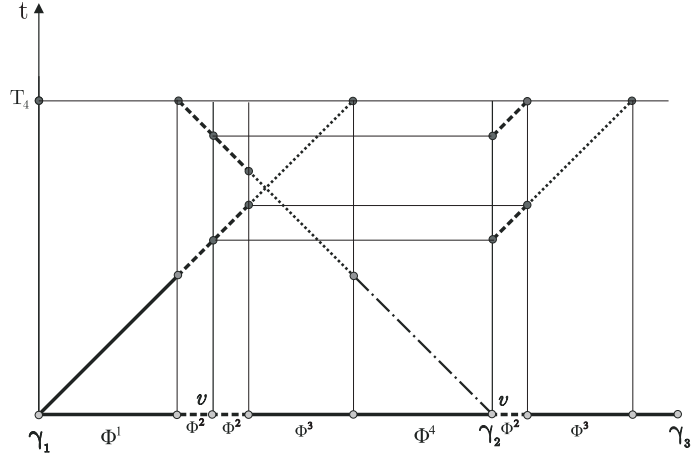}
\caption[Fig.11]{$T=T_4$; the hydras}\label{$T=T_4$; the hydras}
 \end{figure}

$\bullet$\,\,\,Here the partition of the domain
$\Omega^{T_4}[\Sigma]$ is
$\Pi_\Sigma=\Phi^1\cup\Phi^2\cup\Phi^3\cup\Phi^4$: see Fig.
\ref{$T=T_4$; the graph}, \ref{$T=T_4$; the hydras}.

The proper parametrization provides
\smallskip

\noindent$\tau^1_{\gamma_1\,\Phi^1}(r_1)=r_1\!:$\,\,\,
$r_1\in[0,l_1\!+\!l_2\!-\!T_4]$;
\smallskip

\noindent$\tau^1_{\gamma_1\,\Phi^2}(r_2)=l_1-r_2$,\,
$\tau^2_{\gamma_1\,\Phi^2}(r_2)=l_1+r_2\!:$\,\,\,\,$r_2\in[0,T_4-l_2]$;
\smallskip

\noindent$\tau^1_{\gamma_2\,\Phi^2}(r_2)=l_2-r_2$,\,
$\tau^2_{\gamma_2\,\Phi^2}(r_2)=l_2+r_2\!:$\,\,\,\,
$r_2\in[0,T_4-l_2]$;
\smallskip

\noindent$\tau^1_{\gamma_1\,\Phi^3}(r_3)=(l_1\!-\!l_2\!+\!T_4)+r_3$,\,\,
$\tau^1_{\gamma_2\,\Phi^3}(r_3)=(2l_2\!-\!T_4)\!-\!r_3\!:$\,\,\,
$r_3\in[0,l_2-l_1]$;
\smallskip

\noindent$\tau^1_{\gamma_2\,\Phi^4}(r_4)=r_4\!:$\,\,\,\,$r_4\in[0,l_1\!+\!l_2\!-\!T_4]$.
\smallskip

\noindent$\bullet$\,\,\,As one can show, the generators of the
eikonal algebra are of the form
$$
UE^{T_4}_{\gamma_1}U^{-1}=\left(\begin{array}{ccccccc}
r_1  \\
&l_1-r_2&0&0\\
&0& \frac{l_1+r_2}{2} & \frac{l_1+r_2}{2}\\
&0& \frac{l_1+r_2}{2} & \frac{l_1+r_2}{2}\\
&&&&\frac{(l_1-l_2+T_4)+r_3}{2} & \frac{(l_1-l_2+T_4)+r_3}{2}\\
&&&&\frac{(l_1-l_2+T_4)+r_3}{2} & \frac{(l_1-l_2+T_4)+r_3}{2}\\
&&&&&&0
\end{array}\right), $$

$$ UE^{T_4}_{\gamma_2}U^{-1}=\left(\begin{array}{ccccccc}
0 & \\
& \frac{l_2+r_2}{2} & \frac{l_2+r_2}{2}&0\\
& \frac{l_2+r_2}{2} & \frac{l_2+r_2}{2}&0\\
&0&0&l_2-r\\
&&&&0&0\\
&&&&0&(2l_2\!-\!T_4)\!-\!r_3\\
&&&&&& r_4
\end{array}\right)\,.
$$

The blocks $\{(\cdot)_{11}\}$ and $\{(\cdot)_{77}\}$, which
correspond to the families $\Phi^1$ and $\Phi^4$, generate the
com\-mu\-ta\-tive part of the algebra $UE^{T_4}_{\Sigma}U^{-1}$ of
the form $C_0[0,l_1+l_2-T_4]\oplus C_0[0,l_1+l_2-T_4]$.

The blocks $\{(\cdot)_{ij}\}_{i,j=5}^6$ of the generators, which
corres\-pond to the family $\Phi^3$, produce the algebra
$С([0,l_2-l_1];{\mathbb M}^{2})$ in perfect analogy to the moment
$T_3$.

A new type algebra appears owing to the blocks
$\{(\cdot)_{ij}\}_{i,j=2}^4$, which cor\-res\-pond to the family
$\Phi^2$ and are of the form
 \begin{equation}\label{Eq T4 gen1}
\left(\begin{array}{ccccccc}
l_1-r_2&0&0\\
0& \frac{l_1+r_2}{2} & \frac{l_1+r_2}{2}\\
0& \frac{l_1+r_2}{2} & \frac{l_1+r_2}{2}\\
\end{array}\right)=(l_1-r_2)P^1_{\gamma_1\,\Phi^2}+(l_1+r_2)P^2_{\gamma_1\,\Phi^2}
 \end{equation}
and
 \begin{equation}\label{Eq T4 gen2}
\left(\begin{array}{ccccccc}
 \frac{l_2+r_2}{2} & \frac{l_2+r_2}{2}&0\\
 \frac{l_2+r_2}{2} & \frac{l_2+r_2}{2}&0\\
0&0&l_2-r_2\\
\end{array}\right)=(l_2+r_2)P^2_{\gamma_2\,\Phi^2}+(l_2-r_2)P^1_{\gamma_2\,\Phi^2}\,.
 \end{equation}
Further analysis makes use of the properties $1.\!-\!4.$ of the
matrix algebras. First, one can check that
$\vee\{P^i_{\gamma_l\,\Phi^2}\,|\,\,i,l=1,2\}={\mathbb M}^3$.
Then, for $r_2=0$ the generators (\ref{Eq T4 gen1}) and (\ref{Eq
T4 gen2}) turn out to be proportional to the projections
 $$
P_1 :=P^1_{\gamma_1\,\Phi^2}+P^2_{\gamma_1\,\Phi^2}=
\left(\begin{array}{ccccccc}
1&0&0\\
0& \frac{1}{2} & \frac{1}{2}\\
0& \frac{1}{2} & \frac{1}{2}\\
\end{array}\right),\,\,
P_2:=P^2_{\gamma_2\,\Phi^2}+P^1_{\gamma_2\,\Phi^2}=
\left(\begin{array}{ccccccc}
 \frac{1}{2} & \frac{1}{2}&0\\
 \frac{1}{2} & \frac{1}{2}&0\\
0&0&1\\
\end{array}\right),
 $$
which do have the mutual eigenvector $\{1,1,1\}^t\in{\mathbb
R}^3$. Therefore, one easily gets $\vee\{P_1,\,P_2\}\cong{\mathbb
M}^1\oplus{\mathbb M}^2$ that is a proper subalgebra in ${\mathbb
M}^3$. As a result, the contribution of the blocks (\ref{Eq T4
gen1}) and (\ref{Eq T4 gen2}) to the eikonal algebra turns out to
be the standard algebra $\dot{C}([0,T_4-l_2];{\mathbb M}^{3})$.

Also one can verify that there are no more connections at the
endpoints $r=0,\epsilon_\Phi$ between the blocks, which enter in
the generators $UE^{T_4}_{\gamma_1}U^{-1}$ and
$UE^{T_4}_{\gamma_2}U^{-1}$.
\smallskip

\noindent$\bullet$\,\,\,Summarizing, we arrive at the final
representation
 $$
{\mathfrak E}^{T_4}_\Sigma\cong C_0[0,l_1\!+\!l_2\!-\!T_4]\oplus
C_0[0,l_1\!+\!l_2\!-\!T_4]\oplus C([0,l_2\!-\!l_1];{\mathbb
M}^{2})\oplus \dot{C}([0,T_4\!-\!l_2];{\mathbb M}^{3}).
 $$

\end{document}